 \documentclass[12pt, draftclsnofoot, onecolumn]{IEEEtran}

\pdfoutput=1

\usepackage[cmex10]{amsmath}
\usepackage{tikz}
\usepackage{adjustbox}
\tikzset{>=latex}
\usetikzlibrary{calc,shapes,arrows,patterns}
\usepackage{pgfplots} 
\pgfplotsset{compat=1.12}
\usepackage{acronym}

\acrodef{RB}[RB]{resource block}
\acrodef{OSF}[OSF]{oversampling factor}
\acrodef{NF}[NF]{noise figure}
\acrodef{Q-PSK}[Q-PSK]{Quadrature Phase-Shift Keying}
\acrodef{LS}[LS]{least-squares}
\acrodef{AAF}[AAF]{Anti-Aliasing-Filter}
\acrodef{UMTS}[UMTS]{Universal Mobile Telecommunications System}
\acrodef{IIP2}[IIP2]{second-order input intercept point}
\acrodef{MSE}[MSE]{mean-square-error}
\acrodef{NMSE}[NMSE]{normalized mean-square-error}
\acrodef{LMS}[LMS]{least-mean-squares}
\acrodef{RLS}[RLS]{recursive-least-squares}
\acrodef{R-APA}[$\epsilon$-APA]{regularized affine-projection}
\acrodef{NLMS}[$\epsilon$-NLMS]{normalized least-mean-squares}
\acrodef{LMF}[LMF]{least-mean-forth}
\acrodef{NVSSLMS}[$\epsilon$-VSSNLMS]{variable step-size normalized LMS}
\acrodef{NLMF}[NLMF]{normalized least-mean-forth}
\acrodef{pdf}[pdf]{probability density function}
\acrodef{CWGN}[CWGN]{complex white gaussian noise}
\acrodef{CUN}[CUN]{complex uniform noise}
\acrodef{iid}[iid]{independent and identical distributed}
\acrodef{HRM}[HRM]{harmonic rejection mixer}
\acrodef{SNR}[SNR]{signal-to-noise ratio}
\acrodef{LO}[LO]{local oscillator}
\acrodef{CA}[CA]{carrier aggregation}
\acrodef{PWM}[PWM]{pulse width modulation}
\acrodef{PA}[PA]{power amplifier}
\acrodef{Tx}[Tx]{transmit}
\acrodef{Rx}[Rx]{receive}
\acrodef{RMS}[RMS]{root mean square}
\acrodef{3GPP}[3GPP]{3rd Generation Partnership Project}
\acrodef{LNA}[LNA]{low noise amplifier}
\acrodef{PLL}[PLL]{phase-locked loop}
\acrodef{DFE}[DFE]{digital front-end}
\acrodef{LTE}[LTE]{Long Term Evolution}
\acrodef{LTE-A}[LTE-A]{Long Term Evolution-Advanced}
\acrodef{LTE-CA}[LTE-CA]{Long Term Evolution-carrier aggregation}
\acrodef{FDD}[FDD]{frequency division duplex}
\acrodef{TDD}[TDD]{time-division duplex}
\acrodef{ISP}[ISP]{Institute of Signal Processing}
\acrodef{OFDM}[OFDM]{orthogonal frequency-division multiplexing}
\acrodef{UW-OFDM}[UW-OFDM]{Unique Word OFDM}
\acrodef{CMOS}[CMOS]{Complementary Metal-Oxide-Semiconductor}
\acrodef{IM2}[IM2]{second-order intermodulation product}
\acrodef{IMD2}[IMD2]{second-order intermodulation distortion}
\acrodef{IM3}[IM3]{third-order intermodulation product}
\acrodef{FPGA}[FPGA]{Field Programmable Gate Array}
\acrodef{DCRs}[DCRs]{Direct-Conversion Receivers}
\acrodef{DCR}[DCR]{Direct-Conversion Receiver}
\acrodef{OOB}[OOB]{out-of-band}
\acrodef{BP}[BP]{band pass}
\acrodef{BPF}[BPF]{band pass filter}
\acrodef{WPs}[WPs]{work packages}
\acrodef{WP}[WP]{work package}
\acrodef{MOSFETs}[MOSFETs]{Metal Oxide Semiconductor Field Effect Transistors}
\acrodef{RF}[RF]{radio frequency}
\acrodef{RF-DAC}[RF-DAC]{radio frequency digital-to-analog converter}
\acrodef{OFDM}[OFDM]{orthogonal frequency-division multiplexing}
\acrodef{SC-FDMA}[SC-FDMA]{single-carrier frequency-division multiple access}
\acrodef{CW}[CW]{continuous wave}
\acrodef{RM}[RM]{reciprocal mixing}
\acrodef{RAT}[RAT]{radio access technology}
\acrodef{TxL}[TxL]{transmitter leakage}
\acrodef{CSF}[CSF]{channel-select filter}
\acrodef{BB}[BB]{baseband}
\acrodef{FIR}[FIR]{finite impulse response}
\acrodef{IIR}[IIR]{infinite impulse response}
\acrodef{PSD}[PSD]{power spectral density}
\acrodef{SINR}[SINR]{signal-to-interference-plus-noise ratio}
\acrodef{CC}[CC]{component carriers}
\acrodef{CA2}[CA2]{second-order}
\acrodef{CA3}[CA3]{third-order}
\acrodef{UL}[UL]{uplink}
\acrodef{DL}[DL]{downlink}
\acrodef{PCC}[PCC]{primary component carrier}
\acrodef{SCC}[SCC]{secondary component carrier}
\acrodef{VSWR}[VSWR]{voltage-standing-wave-ratio}
\acrodef{ADC}[ADC]{analog-to-digital converter}
\acrodef{DC}[DC]{direct-current}
\acrodef{PN}[PN]{phase noise}
\acrodef{HR}[HR]{harmonic rejection}
\acrodef{DCO}[DCO]{digitally controlled oscillator}

\usepackage{amssymb}
\usepackage{hyperref}
\hypersetup{pageanchor=false} 
\hypersetup{
    colorlinks=false,
    pdfborder={0 0 0},
}
\usepackage[nospace]{cite}
\usepackage{multirow}

\newcommand{\bm}{\mathbf}

\title{\LARGE \bf
A Robust Nonlinear RLS Type Adaptive Filter for Second-Order-Intermodulation Distortion Cancellation in FDD LTE and 5G Direct Conversion Transceivers 
}

\author{
	\IEEEauthorblockN{Andreas Gebhard\IEEEauthorrefmark{1}, Oliver Lang\IEEEauthorrefmark{3}, Michael Lunglmayr\IEEEauthorrefmark{3}, Christian Motz\IEEEauthorrefmark{1} \\ Ram Sunil Kanumalli\IEEEauthorrefmark{2},  Christina Auer\IEEEauthorrefmark{1}, Thomas Paireder\IEEEauthorrefmark{1}, Matthias Wagner\IEEEauthorrefmark{1}, Harald Pretl\IEEEauthorrefmark{4}\IEEEauthorrefmark{2}
and Mario Huemer\IEEEauthorrefmark{1}}
	\IEEEauthorblockA{\IEEEauthorrefmark{1}
	Christian Doppler Laboratory for Digitally Assisted RF Transceivers for Future Mobile Communications,\\ Institute of Signal Processing, Johannes Kepler University, Linz, Austria}
		\IEEEauthorblockA{\IEEEauthorrefmark{3}Institute of Signal Processing, Johannes Kepler University, Linz, Austria}
	\IEEEauthorblockA{\IEEEauthorrefmark{2}Danube Mobile Communications Engineering GmbH \& Co KG, Freist\a"adter Stra{\ss}e 400, 4040 Linz, Austria}
		\IEEEauthorblockA{\IEEEauthorrefmark{4}
	Institute for Integrated Circuits, Johannes Kepler University, Linz, Austria}
	{Email: andreas.gebhard@jku.at}
}
\providecommand{\keywords}[1]{\textbf{\textit{Index terms ---}} #1}

\newcommand{\vx}{\ve{x}}

\begin{document}

\tikzstyle{block} = [draw, fill=blue!20, rectangle, minimum height=3em, minimum width=1em]
\tikzstyle{sum} = [draw, circle, node distance=1cm]
\tikzstyle{input} = [coordinate]
\tikzstyle{output} = [coordinate]
\tikzstyle{pinstyle} = [pin edge={to-,thin,black}]

\definecolor{red}{rgb}{1,0,0}

\setlength{\textfloatsep}{5pt} 

\maketitle
\thispagestyle{empty}
\pagestyle{empty}

\setlength{\textfloatsep}{5pt} 
\newcommand{\ve}{\mathbf}
\newcommand{\m}{\mathbf}
\newcommand{\veit}[1]{\mathitbf{#1}} 
\newcommand{\veel}[1]{#1} 

\newcommand{\mf}[1]{\mathbf{\tilde{\mathbf{#1}}}} 
\newcommand{\vef}[1]{\mathbf{\tilde{\mathbf{#1}}}} 
\newcommand{\vefel}[1]{\tilde{#1}} 
\newcommand{\mfit}[1]{\mathitbf{\tilde{\mathitbf{#1}}}} 
\newcommand{\vefit}[1]{\mathitbf{\tilde{\mathitbf{#1}}}} 

\begin{abstract} 
\acresetall
Transceivers operating in frequency division duplex experience a \ac{TxL} signal into the receiver due to the limited duplexer stop-band isolation. This \ac{TxL} signal in combination with the second-order nonlinearity of the receive mixer may lead to a \ac{BB} \ac{IMD2} with twice the transmit signal bandwidth. In direct conversion receivers, this nonlinear \ac{IMD2} interference may cause a severe signal-to-interference-plus-noise ratio degradation of the wanted receive signal. 
This contribution presents a nonlinear Wiener model \ac{RLS} type adaptive filter for the cancellation of the \ac{IMD2} interference in the digital \ac{BB}. The included \mbox{channel-select-,} and DC-notch filter at the output of the proposed adaptive filter ensure that the provided \ac{IMD2} replica includes the receiver front-end filtering.
%
%
A second, robust version of the nonlinear \ac{RLS} algorithm is derived which provides numerical stability for highly correlated input signals which arise in e.g. LTE-A intra-band multi-cluster transmission scenarios. The performance of the proposed algorithms is evaluated by numerical simulations and by measurement data. 

\acresetall
\end{abstract}
\section{Introduction} 
\acresetall
Modern \ac{RF} transceivers are enhanced by digital signal processing to mitigate non-idealities in the analog front-end. One of the main reasons of receiver desensitization in \ac{FDD} transceivers is the limited duplexer isolation between the transmitter and the receiver which is around 50\,dB to 55\,dB \cite{Ericsson_1,Vazny_1}. The resulting \ac{TxL} signal can be identified as the root cause of several receiver \ac{BB} interferences. Especially in \ac{CA} receivers multiple clock sources are needed to cover the different \ac{CA} scenarios and band combinations. Due to cross-talk between the receivers on the chip and device nonlinearities, spurs appear in the receiver front-end. 
\begin{figure}[ht]
\centering
\begin{tikzpicture}
\node (img) at (0,0) {\includegraphics[width=1\columnwidth,trim=0cm 0.25cm 0cm 0.25cm]{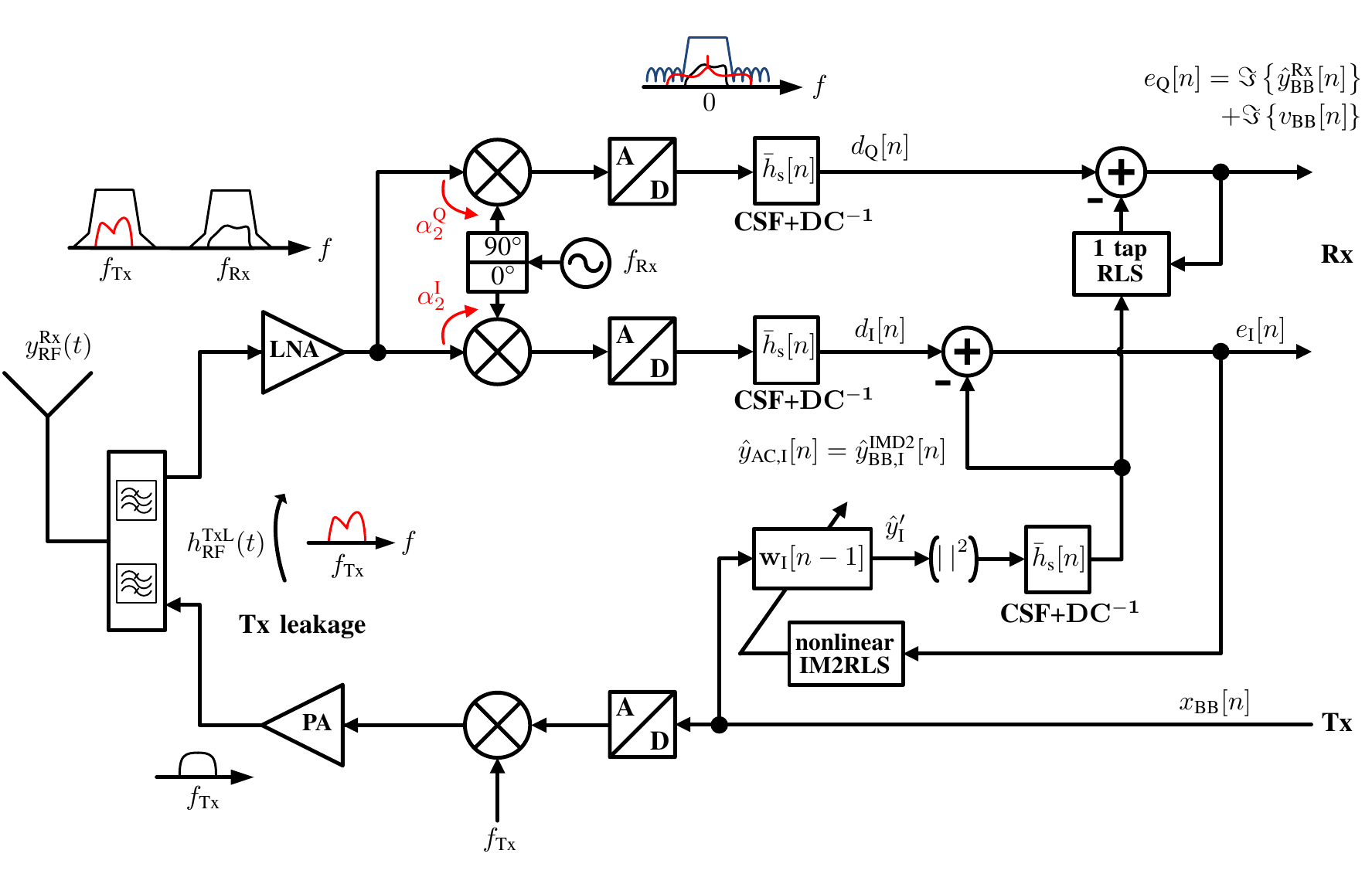}};
\end{tikzpicture}
\caption{
Block diagram depicting an \ac{RF} transceiver operating in \ac{FDD} mode which experiences a second-order intermodulation distortion in the receiver due to the transmitter leakage signal and the \ac{Rx} mixer \ac{RF}-to-\ac{LO} terminal coupling. A nonlinear \ac{RLS}-type adaptive filters is used to estimate the \mbox{I-path} IMD2 interference. The Q-path IMD2 interference is estimated with a linear 1-tap RLS adaptive filter which uses the estimated I-path IMD2 replica as reference input.
}
\label{fig:IMD2_block_diagram}
\end{figure}

If such a spur falls near the actual \ac{Tx} frequency, then the \ac{TxL} signal is down-converted into the \ac{Rx} \ac{BB} where it causes a \ac{SINR} degradation of the wanted receive signal. The cancellation of this so called modulated spurs with adaptive filtering is demonstrated in \cite{Gebhard2016,Kanumalli2016}. 

Another prominent interference caused by the \ac{TxL} signal and the second-order nonlinearity of the receiver is the \ac{IMD2}. This second-order nonlinear distortion is caused by e.g. a coupling between the \ac{RF}- and \ac{LO}-ports in the I-, and Q-path of the \ac{Rx} IQ-mixer as indicated in Fig.~\ref{fig:IMD2_block_diagram} \cite{Razavi_1}. 
An interesting fact of this nonlinear interference is, that one part of the generated second-order intermodulation products always falls around zero-frequency independent of the \ac{Tx}-to-\ac{Rx} frequency offset (duplexing distance). In case of direct-conversion receiver architectures, this leads to a degradation of the wanted receive signal.

The mathematical modeling in \cite{Gebhard2017,Kiayani_1} shows that the \ac{BB} \ac{IMD2} interference contains the squared envelope of the \ac{BB} equivalent \ac{TxL} signal. The resulting \ac{BB} \ac{IMD2} interference has twice the \ac{Tx} signal bandwidth and contains a DC due to the envelope-squaring. In the receiver front-end, the overall DC arising from a number of sources is canceled by a mixed-signal cancellation to prevent the \ac{ADC} from saturation. In the digital domain, the signal is filtered by a \ac{CSF} to reduce its bandwidth to the \ac{LTE} signal bandwidth. 

In the existing literature, the authors of \cite{Lederer_1,Frotzscher_3,Kahrizi_1} discussed adaptive \ac{LMS} type \ac{IMD2} interference cancellation algorithms for frequency-flat duplexer stop-bands. In \cite{Frotzscher_2} a Volterra kernel based \ac{LS} approach for frequency-selective Tx-Rx responses is proposed. The authors in \cite{Kiayani_1} presented a two-step \ac{LS} approach for the \ac{IMD2} cancellation and considered a static 3rd-order \ac{PA} nonlinearity and IQ-imbalance in the transmit mixer. In \cite{Gheidi_1} a \ac{Tx} \ac{CA} transceiver is considered where the transmit signal of both transmitters leaks through a diplexer into one unpaired \ac{CA} receiver. The diplexer stop-band is modeled as a first-order \ac{FIR} system which states a nearly frequency-flat response. The authors incorporated a fourth-order nonlinearity without memory into the estimation process, which results in an \ac{LS} problem with four unknown coefficients.  

This contribution presents a nonlinear Wiener model \ac{RLS} type adaptive filter (IM2RLS) with exponential forgetting factor which is suitable for highly frequency selective duplexer stop-band frequency responses like indicated in Fig. \ref{fig:spectrum}. It targets the digital \ac{IMD2} cancellation for high performance cellular base stations and mobile phones. 
The Wiener model uses a static nonlinearity at the output of the adaptive filter which has the advantage that less coefficients are needed in the estimation process compared to a Volterra kernel based adaptive filter \cite{mathews2000polynomial}.

An additional version of the proposed algorithm is presented which enhances the algorithm by a DC-notch filter to cancel the DC in the interference replica. This is needed because direct-conversion receivers employ a DC cancellation to suppress the DC in order to prevent the \ac{ADC} from saturation. The DC in the received signal is time-variant and has many sources like e.g. LO-LO self mixing \cite{Razavi_1}, and therefore must not be related explicitly to the DC which is generated by the \ac{IMD2} interference.
Consequently, the \ac{IMD2} interference related DC is removed from the received signal which complicates the \ac{IMD2} replica estimation. This DC removal is considered in \cite{Frotzscher_2,Gebhard2017}, and neglected in \cite{Lederer_1,Lederer_3,Frotzscher_3,Kiayani_1}.

The derived IM2RLS with DC-notch filter is extended by a regularization \mbox{(R-IM2RLS)} which makes the algorithm applicable for highly correlated \ac{BB} transmit signals where the autocorrelation matrix can be close to singular. A high correlation in the transmit signal can be due to oversampling which happens e.g. in the case of multi-cluster transmissions (introduced in 3GPP LTE-A Release 11) where only a part of the available \acp{RB} are allocated. The presented IM2RLS algorithm is an extension to the nonlinear \ac{LMS} type adaptive filter derived in \cite{Gebhard2017} with improved steady-state cancellation and convergence speed. 

The structure of the presented work is as follows: Section \ref{sec:problem_statement} explains the \ac{IIP2} characterization and demonstrates the degradation of the \ac{Rx} performance due to the \ac{IMD2} interference. Section \ref{sec:system_model} provides a detailed \ac{IMD2} interference model which motivates the proposed structure of the nonlinear adaptive filter. In Section \ref{sec:IM2RLS}, the IM2RLS algorithm is derived and the impact of adding a DC-notch filter to the algorithm is evaluated. The R-IM2RLS alrorithm is derived in section \ref{sec:R-IM2RLS} which is robust against highly correlated input signals as they occur in intra-band multi-cluster transmissions. Finally, in the sections \ref{sec:simulations} and \ref{sec:measurements}, the performance of the R-IM2RLS algorithm is evaluated with simulations and measured data using RF components.
\section{Problem statement} \label{sec:problem_statement}
The receiver \ac{IIP2} is characterized by using two cosine signals with the frequencies $f_1$ and $f_2$ of equal amplitude and the total power $P_{\text{in,2t}}$ at the input of the nonlinear mixer. The resulting total \ac{IMD2} power generated at DC, $f_1+f_2$ and $f_2-f_1$ at the output of the mixer can be calculated by \mbox{$P_{\text{IM2}}^{\text{Tot,2t}}=2 P_{\text{in,2t}}-\text{IIP2}_{\text{2t}}$} \cite{liu2009ip2}, where $\text{IIP2}$ is the two-tone IIP2 value in dBm. Here, half of the total \ac{IMD2} power falls to DC, and one quarter each to $f_1+f_2$ and $f_2-f_1$. To characterize the \ac{IIP2} in a zero-IF receiver, the frequencies $f_1$ and $f_2$ are chosen such that $f_2-f_1$ falls within the \ac{CSF} bandwidth. Thereby the power at $f_2-f_1$ is measured and the \ac{IIP2} is determined by \mbox{$\text{IIP2}_{\text{2t}}=2 P_{\text{in,2t}}-P_{\text{IM2}}^{f_2-f_1}-6\,\text{dB}$}. 

For modulated signals, the \ac{BB} \ac{IMD2} power is modulation dependent and further reduced by the \ac{CSF}. This is considered by a correction-factor which corrects the \ac{IMD2} power calculated by the two-tone formula \cite{Walid_1,Atalla_1}.

Although the DC-, and channel-select filtering in the receiver reduces the \ac{IMD2} \ac{BB} interference power by 6\,dB in the two-tone signal case \cite{liu2009ip2}, and by about 13.4\,dB \cite{Gebhard2017,Walid_1,Atalla_1} in the case of modulated \ac{Tx} signals, the left-over \ac{IMD2} interference may lead to a severe \ac{SNR} degradation of the wanted \ac{Rx} signal in reference sensitivity cases \cite{3GPP_2}. Assuming a transmitter power of 23\,dBm at the antenna, and an average \ac{Tx}-to-\ac{Rx} duplexer isolation at the transmit frequency of 50\,dB, the \ac{TxL} signal power at the input of the receiver is \mbox{$P_{\text{RF}}^{\text{TxL}}=23\,\text{dBm}-50\,\text{dB} =-27\,\text{dBm}$}. After amplification with the \ac{LNA} gain which is assumed as 20\,dB, the \ac{RF} \ac{TxL} signal power increases to \mbox{$P_{\text{RF}}^{\text{TxL}}=-7\,\text{dBm}$} at the input of the nonlinear mixer. 

\begin{figure}[h]
\centering
\begin{tikzpicture}
	\begin{axis}[
	    legend style={font=\scriptsize},
			width=0.6*\columnwidth, 
			height = .45\columnwidth,
			legend columns = {2},
			grid, 
			xlabel={$f$ [MHz]}, 
			ylabel={PSD [dBm/15\,kHz]},
			xmin = -8,
			xmax = 8,
			ymin = -160,
			ymax = -20,
			ytick={-160,-140,-120,...,-60,-40,-20},
			legend style={at={(0.295,0.01)},anchor=south west}
			]
			 
			
			\addplot[color=black, thick, mark = x, mark repeat={5}] table[x index =0, y index =3] {plot_data/spectrum_full_allocation_IIP2_50dBm.dat};
			\addlegendentry{$y_{\text{BB}}^{\text{TxL}}$}
			
			
			\addplot[color=red, thick] table[x index =0, y index =1] {plot_data/spectrum_full_allocation_IIP2_50dBm.dat};
			\addlegendentry{$y_{\text{BB}}^{\text{Tot}}$}
			
			\addplot[color=green, thick, style = dashed] table[x index =0, y index =2] {plot_data/spectrum_full_allocation.dat};
			\addlegendentry{$y_{\text{BB}}^{\text{Rx}}$}
			
						\addplot[color=blue, thick] table[x index =0, y index =4] {plot_data/spectrum_full_allocation_IIP2_50dBm.dat};
			\addlegendentry{$y_{\text{BB}}^{\text{IMD2}}$}
			
			\addplot[color=black, thick, style = densely dotted] table[x index =0, y index =6] {plot_data/spectrum_full_allocation_IIP2_50dBm.dat};
			\addlegendentry{Noise}
			
  \end{axis}
\end{tikzpicture}
\caption{Equivalent \ac{BB} spectrum of the frequency-selective \ac{Tx} leakage signal $y_{\text{BB}}^{\text{TxL}}$ (the corresponding passband signal is located at $f_{\text{Tx}}$) and the total received signal $y_{\text{BB}}^{\text{Tot}}$ after amplification with 20\,dB \ac{LNA} gain. The wanted Rx signal with SNR\,=\,3\,dB, and the receiver noise floor after amplification with 20\,dB \ac{LNA} gain are at -77\,dBm and \mbox{-80\,dBm}\,$\widehat{=}$\,\mbox{-108.2\,dBm/15\,kHz} respectively. The total received signal contains the DC-, and channel-select filtered \ac{IMD2} interference with $P_{\text{Tx}}$\,=\,23\,dBm at an assumed IIP2 of 50\,dBm.
}
\label{fig:spectrum}
\end{figure}
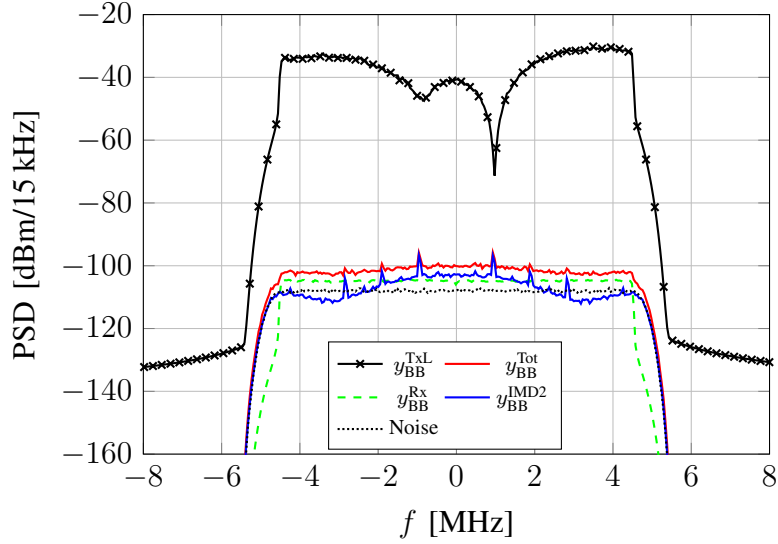

The two-tone \ac{IIP2} value of typical \ac{RF} mixers is between 50\,dBm and 70\,dBm \cite{Madadi_1,Dufrene_1}. Assuming an \ac{IIP2} of 60\,dBm, the resulting \ac{BB} \ac{IMD2} power with a full allocated LTE10 QPSK modulated transmission and the determined correction factor of \mbox{$\text{CF} = 13.4\,\text{dB}$} is \\ \mbox{$P_{\text{IM2}}^{\text{CSF,LTE}}= 2 P_{\text{RF}}^{\text{TxL}} - \text{IIP2} - \text{CF} = -87.4 \, \text{dBm}$} ~\cite{Gebhard2017}. In an LTE10 reference sensitivity case, the wanted signal power at the antenna can be as low as -97\,dBm \cite{3GPP_2}. The thermal noise power within 10\,MHz bandwidth is -104.5\,dBm and the assumed receiver \ac{NF} is 4.5\,dB which results in a receiver noise floor at -100\,dBm. After amplification with 20\,dB \ac{LNA} gain, the wanted signal power is -77\,dBm and the noise floor at -80\,dBm corresponding to an \ac{Rx} \ac{SNR} of 3\,dB. The \ac{SNR} drops from 3\,dB to an \ac{SINR} of 2.27\,dB due to the \ac{IMD2} interference assuming an \ac{IIP2} of +60\,dBm. In case of an reduced IIP2 of 55\,dBm / 50\,dBm, the \ac{SINR} drops even further to 1\,dB / -1.4\,dB, respectively. Fig.~\ref{fig:spectrum} depicts the spectrum of the frequency selective \ac{BB} equivalent \ac{TxL} signal $y_{\text{BB}}^{\text{TxL}}$ which generates the complex valued \ac{IMD2} interference $y_{\text{BB}}^{\text{IMD2}}$ by a coupling between the RF-to-LO terminals of the I-, and Q-path mixer. The total received signal $y_{\text{BB}}^{\text{Tot}}$ contains the wanted \ac{Rx} signal $y_{\text{BB}}^{\text{Rx}}$ which is degraded by the \ac{IMD2} interference and the noise.   
\section{System model}\label{sec:system_model}  
\subsection{\ac{IMD2} Interference Model}
Based on the block diagram in Fig.~\ref{fig:IMD2_block_diagram} depicting an \ac{RF} transceiver operating in \ac{FDD} mode, a detailed \ac{IMD2} interference model is derived. The used mathematical operators $\left(.\right)^*$, $\left(.\right)^T$, $\left(.\right)^H$, and $\ast$ denote the complex conjugate, transpose, Hermitian transpose, and convolution, respectively. The complex \ac{BB} transmit signal \mbox{$x_{\text{BB}}(t)=x_{\text{I}}(t)+jx_{\text{Q}}(t)$} is up-converted to the passband and amplified by the linearly assumed \ac{PA} with gain $A_{\text{PA}}$ resulting in the \ac{RF} transmit signal 
\begin{equation}
x_{\text{RF}}(t) = A_{\text{PA}}\Re\left\{x_{\text{BB}}(t)e^{j2\pi f_{\text{Tx}}t}\right\}.
\end{equation}
This signal leaks through the duplexer \ac{RF} stop-band impulse response
\begin{equation}
h_{\text{RF}}^{\text{TxL}}(t)=2\Re\left\{h_{\text{BB}}^{\text{TxL}}(t) e^{j2\pi f_{\text{Tx}}t}\right\},
\end{equation}
which is modeled by the \ac{BB} equivalent duplexer impulse response $h_{\text{BB}}^{\text{TxL}}(t)$ into the receiver, thereby creating the \ac{TxL} signal
\begin{equation}\label{eq:y_RF_TxL}
\begin{aligned}
y_{\text{RF}}^{\text{TxL}}(t) &= x_{\text{RF}}(t) \ast h_{\text{RF}}^{\text{TxL}}(t)\\
&=A_{\text{PA}} \Re\left\{\left[x_{\text{BB}}(t) \ast h_{\text{BB}}^{\text{TxL}}(t)\right]e^{j2\pi f_{\text{Tx}}t}\right\}.
\end{aligned}
\end{equation}
The received signal at the output of the \ac{LNA} with gain $A_{\text{LNA}}$ 
\begin{equation}\label{eq:y_Tot_after_LNA}
y_{\text{RF,LNA}}^{\text{Tot}}(t)=A_{\text{LNA}}\left[y_{\text{RF}}^{\text{TxL}}(t)+y_{\text{RF}}^{\text{Rx}}(t)+v_{\text{RF}}(t)\right],
\end{equation}
is composed by the amplified \ac{TxL} signal, the wanted Rx signal $y_{\text{RF}}^{\text{Rx}}(t)$ and the noise signal $v_{\text{RF}}(t)$.
The output signal of the I-, and Q-path mixer is combined into the complex valued signal $y_{\text{RF,mixer}}^{\text{Tot}}(t)$ \eqref{eq:mix_out_sig}. It contains the wanted signal which is down-converted with the linear gain \mbox{$\alpha_1 = \alpha_1^{\text{I}}+ j \alpha_1^{\text{Q}}$}, and the second order interference with the mixer \ac{RF}-to-\ac{LO} terminal coupling coefficient $\alpha_2=\alpha_2^\text{I} + j \alpha_2^\text{Q}$.
\begin{equation}\label{eq:mix_out_sig}
\begin{aligned}
y_{\text{RF,mixer}}^{\text{Tot}}(t)&= y_{\text{RF,LNA}}^{\text{Tot}}(t)\alpha_1^{\text{I}} cos\left(2\pi f_{\text{Rx}}t\right) \\
& + y_{\text{RF,LNA}}^{\text{Tot}}(t) \left[\alpha_2^{\text{I}} y_{\text{RF,LNA}}^{\text{Tot}}(t)\right]\\
& - j y_{\text{RF,LNA}}^{\text{Tot}}(t)\alpha_1^{\text{Q}} sin\left(2\pi f_{\text{Rx}}t\right)\\
& + j y_{\text{RF,LNA}}^{\text{Tot}}(t) \left[\alpha_2^{\text{Q}} y_{\text{RF,LNA}}^{\text{Tot}}(t)\right] \\
& = y_{\text{RF,LNA}}^{\text{Tot}}(t)\alpha_1 e^{-j2\pi f_{\text{Rx}}t} + \alpha_2 \, y_{\text{RF,LNA}}^{\text{Tot}}(t)^2
\end{aligned}
\end{equation}
%
%
Assuming a direct conversion receiver, and using the identity \mbox{$\Re\left\{\eta e^{j\kappa}\right\}=\frac{1}{2}\left(\eta e^{j\kappa}+\eta^*e^{-j\kappa}\right)$}, the total mixer output signal by neglecting the signal content which falls outside the \ac{BB} bandwidth becomes
\begin{equation}\label{eq:y_Tot_after_mixer}
\begin{aligned}
y_{\text{RF,mixer}}^{\text{Tot}}(t)&=\alpha_1\frac{A_{\text{LNA}}}{2} y_{\text{BB}}^{\text{Rx}}(t) +\alpha_1 \frac{A_{\text{LNA}}}{2} v_{\text{BB}}(t)\\
&+  \frac{\alpha_2}{2}
\cdot \left(\left|A_{\text{LNA}} A_{\text{PA}} \, x_{\text{BB}}(t)*h_{\text{BB}}^{\text{TxL}}(t) \right|^2 +\frac{1}{2}\left|y_{\text{BB}}^{\text{Rx}}(t)\right|^2 \right. \\
&\left. + \Re\left\{y_{\text{BB}}^{\text{Rx}}(t) v^*_{\text{BB}}(t)  \right\} +\frac{1}{2}\left|v_{\text{BB}}(t)\right|^2 \right).
\end{aligned}
\end{equation}
As $\left|\alpha_2\right|<<1$, the three last terms in \eqref{eq:y_Tot_after_mixer} may be neglected \cite{Kiayani_1,Gebhard2017}. The total received discrete-time \ac{BB} signal including the DC-cancellation and channel-select filtering becomes  
%
\begin{equation}\label{eq:y_Tot_after_CSF}
\begin{aligned}
y_{\text{BB}}^{\text{Tot}}[n]&= \alpha_1 \frac{A_{\text{LNA}}}{2} y_{\text{BB}}^{\text{Rx}}[n] \ast \bar{h}_{\text{s}}[n] + \alpha_1 \frac{A_{\text{LNA}}}{2} v_{\text{BB}}[n] \ast \bar{h}_{\text{s}}[n]\\
&+ \underbrace{\frac{\alpha_2}{2}\left| A_{\text{LNA}} A_{\text{PA}} x_{\text{BB}}[n]*h_{\text{BB}}^{\text{TxL}}[n] \right|^2  \ast \bar{h}_{\text{s}}[n]}_{y_{\text{BB}}^{\text{IMD2}}[n]},
\end{aligned}
\end{equation}
where the DC-, and \ac{CSF} are combined in the impulse response \mbox{$\bar{h}_s[n] = h_{\text{DC}}[n] \ast h_s[n]$}. Here, \mbox{$h_{\text{BB}}^{\text{TxL}}[n]=T_s h_{\text{BB}}^{\text{TxL}}(t)\big|_{t=nT_s}$} is the impulse invariant \mbox{\cite{oppenheim2010discrete,tzschoppe2009causal}}, scaled and sampled version of the continuous-time \ac{BB} duplexer impulse response $h_{\text{BB}}^{\text{TxL}}(t)$.
\subsection{Interference Replica Model}
For the adaptive filter development to cancel the \ac{IMD2} interference in the digital \ac{BB}, the interference model \eqref{eq:y_Tot_after_CSF} is rewritten to the form
\begin{equation}\label{eq:y_Tot_after_CSF_adaptive_filter}
\begin{aligned}
y_{\text{BB}}^{\text{Tot}}[n]&= \underbrace{\frac{\alpha_2^{\text{I}}}{2}\left| A_{\text{LNA}} A_{\text{PA}} x_{\text{BB}}[n]*h_{\text{BB}}^{\text{TxL}}[n] \right|^2  \ast \bar{h}_{\text{s}}[n]}_{y_{\text{BB}}^{\text{IMD2,I}}[n]} \\
&+j \underbrace{\frac{\alpha_2^{\text{Q}}}{2}\left| A_{\text{LNA}} A_{\text{PA}} x_{\text{BB}}[n]*h_{\text{BB}}^{\text{TxL}}[n] \right|^2  \ast \bar{h}_{\text{s}}[n]}_{y_{\text{BB}}^{\text{IMD2,Q}}[n]} + v_{\text{BB}}'[n]
\end{aligned}
\end{equation}
where the complex valued wanted signal and the noise signal are combined in $v_{\text{BB}}'[n]$.
Assuming $\alpha_2^{\text{I}} > 0$, and approximating the duplexer impulse response $h_{\text{BB}}^{\text{TxL}}[n]$ by the \ac{FIR} impulse response vector $\ve{h}_{\text{BB}}^{\text{TxL}}$ of length $N_{\ve{w}}$, we can rewrite the model \eqref{eq:y_Tot_after_CSF_adaptive_filter} further to  
\begin{equation}\label{eq:final_IMD2_model}
\begin{aligned}
y_{\text{BB}}^{\text{Tot}}[n]&= \left|\ve{x}^T[n] \ve{h}_{\text{I}} \right|^2 \ast \bar{h}_{\text{s}}[n] +j \left|\ve{x}^T[n] \ve{h}_{\text{Q}} \right|^2 \ast \bar{h}_{\text{s}}[n] + v_{\text{BB}}'[n] \\
&=y_{\text{BB}}^{\text{IMD2,I}}[n] + j \epsilon \, y_{\text{BB}}^{\text{IMD2,I}}[n] + v_{\text{BB}}'[n],
\end{aligned}
\end{equation}
where $\ve{h}_{\text{I}}$ and $\ve{h}_{\text{Q}}$ are incorporating $\ve{h}_{\text{BB}}^{\text{TxL}}$ and all scalar scaling factors in the I-, and Q-path respectively. The used vector $\ve{x}[n]$ is the complex valued tapped delay-line input signal vector \mbox{$\ve{x}[n]=\left[x_{\text{BB}}[n],x_{\text{BB}}[n-1],\ldots,x_{\text{BB}}[n-N_{\ve{w}}+1]\right]^T$}, and the real valued scaling factor $\epsilon$ shows that the Q-path \ac{IMD2} interference may be modeled as a scaled version of the \mbox{I-path} interference. Motivated by the model \eqref{eq:final_IMD2_model} we propose the I-path \ac{IMD2} interference replica model 
\begin{equation}\label{eq:IMD2_model}
\begin{aligned}
\hat{y}_{\text{AC,I}}[n]= \left|\ve{x}^T[n] \ve{w}_{\text{I}}[n] \right|^2  \ast \bar{h}_\text{s}[n],
\end{aligned}
\end{equation}
using the adaptive filter coefficient vector $\ve{w}_{\text{I}}[n]$. The index AC indicates the DC cancellation in the \ac{IMD2} replica generation. The replica model comprises an adaptive Wiener model \ac{FIR} filter where the output signal is DC-, and channel-select filtered. The Q-path \ac{IMD2} interference is generated by estimating the scaling parameter $\epsilon$ by a linear single-tap \ac{RLS} algorithm which uses the estimated I-path \ac{IMD2} interference as reference input. This model is used to derive the adaptive filter structure shown in Fig. \ref{fig:IMD2_block_diagram} to cancel the \ac{IMD2} interference in the digital \ac{BB}. For the case if $\alpha_2^{\text{I}} < 0$, the sign of the desired signal in the I-path $d_{\text{I}}$ and the replica signal of the adaptive filter need to be changed.

%
\section{Nonlinear Recursive Least-Squares algorithm}\label{sec:IM2RLS}
\begin{figure}[ht]
\centering
\input{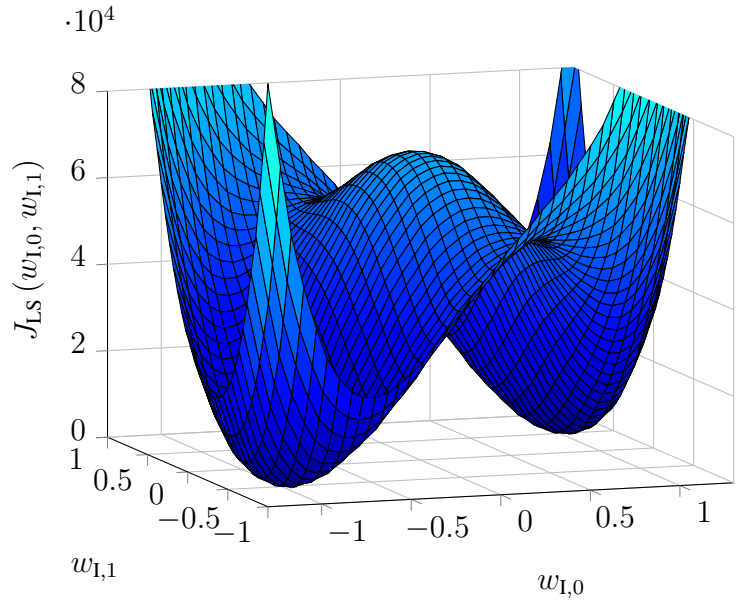}
\caption{Shape of the cost function \eqref{eq:LS_cost_function_sum} for white Gaussian input signals with $\lambda=1$ and for the real valued coefficient vector \mbox{$\ve{h}_{\text{I}}=[1, 0.5]^T$} when the desired signal $d_{\text{I}}[n]$ and the \ac{IMD2} replica are containing the DC. At the origin $\ve{w}_{\text{I}}=\bm{0}$, a local maximum can be observed.}
\label{fig:LS_cost_function_3d_with_DC}
\end{figure}

In this section, a nonlinear Wiener model \ac{RLS} type adaptive filter to estimate the channel-select filtered I-path \ac{IMD2} interference is developed. In a first step the IM2RLS algorithm without DC-notch filter, which implies that the received signal contains the DC, is developed. Therefore, the replica model \eqref{eq:IMD2_model} without DC cancellation 
\begin{equation}\label{eq:I_patch_IMD2_replica}
\begin{aligned}
\hat{y}_{\text{I}}[n] &= \left|\ve{x}^T[n] \ve{w}_{\text{I}}[n] \right|^2 \ast h_\text{s}[n] \\
& = \ve{x}^T[n] \ve{w}_{\text{I}}[n] \ve{x}^H[n] \ve{w}_{\text{I}}^*[n] \ast h_\text{s}[n] 
\end{aligned}
\end{equation}
%
is used. The \ac{LS} cost function up to the time index $n$ with the exponential forgetting factor \mbox{$0<<\lambda\leq1$} is
\begin{equation}\label{eq:LS_cost_function_sum}
\begin{aligned}
J_{\text{LS}}[n] &= \sum^{n}_{i=0}\lambda^{n-i}\left|d_{\text{I}}[i] - \ve{x}^T[i] \ve{w}_{\text{I}}[n] \ve{x}^H[i] \ve{w}_{\text{I}}^*[n] \ast h_\text{s}[i]\right|^2.
\end{aligned}
\end{equation}
This cost function is visualized in Fig. \ref{fig:LS_cost_function_3d_with_DC} for an example impulse response $\ve{h}_{\text{I}}=\left[1, 0.5\right]^T$ and $\lambda = 1$ where the estimated coefficients $w_{\text{I,0}}$ and $w_{\text{I,1}}$ are constrained to be real valued. Two equivalent global minimum points and a local maximum at the origin $\ve{w}_{\text{I}}=\bm{0}$ can be observed. The two solutions \mbox{$\ve{w}_{\text{I,1}}=[1, 0.5]^T$}, and \mbox{$\ve{w}_{\text{I,2}} = [-1, -0.5]^T$} minimize the cost function which can be explained with the absolute-squaring nature of the \ac{IMD2} interference. Both solutions lead to the same \ac{IMD2} replica signal.
%
%
%
%
%
Assuming real valued \ac{CSF} impulse response coefficients $h_s[n]$, and observing that $d_{\text{I}}[i]$ is the desired signal in the I-path, and therefore real valued, the gradient of the cost function \eqref{eq:LS_cost_function_sum} may be derived. The gradient of the cost function with respect to the conjugate coefficient vector $\ve{w}^*_{\text{I}}$ using the Wirtinger calculus \cite{Brandwood_1,Van_den_Bos_1,mandic2009complex} becomes
\begin{equation}\label{eq:RLS1_gradient}
\begin{aligned}
\nabla_{\ve{w}^*_{\text{I}}} J_{\text{LS}}&= \left[\frac{\partial J_{\text{LS}}[n]}{\partial \ve{w}^*_{\text{I}}[n]}\right]^T\\
&=\sum^{n}_{i=0}\lambda^{n-i} \left[-2\, d_{\text{I}}[i]\vx^T[i] \ve{w}_{\text{I}}[n] \vx^*[i] \ast h_s[i] \right.\\
&\left. \hspace{2cm} +2 \left(\vx^T[i] \ve{w}_{\text{I}}[n] \vx^*[i] \ast h_s[i]\right) \right.\\
& \left. \hspace{2cm} \cdot \left(\vx^H[i] \ve{w}_{\text{I}}^*[n] \vx^T[i] \ast h_s[i]\right) \ve{w}_{\text{I}}[n] \right].
\end{aligned}
\end{equation}
By setting the gradient to zero, the Wiener Filter equation is obtained by
\begin{equation}\label{eq:RLS1_wiener_equation}
\begin{aligned}
\tilde{\bm{R}}\left(\ve{w}_{\text{I}}[n]\right)\ve{w}_{\text{I}}[n]&=\tilde{\ve{r}}\left(\ve{w}_{\text{I}}[n]\right),
\end{aligned}
\end{equation} 
where it can be observed that the autocorrelation matrix $\tilde{\bm{R}}$ and the cross-correlation vector $\tilde{\ve{r}}$ are functions of the unknown coefficient vector $\ve{w}_{\text{I}}[n]$. In a slowly varying or nearly stationary system environment it can be assumed that $\ve{x}^T[i] \ve{w}[n]\approx \ve{x}^T[i] \ve{w}[i-1]$ when the index $i$ is close to $n$ \cite{Chen_CMA_1,Chen_CMA_2}. If the index $i << n$, the approximation introduces an error which is however attenuated by the forgetting factor. Defining the new cost function 
\begin{equation}\label{eq:LS_cost_function_sum_new}
\begin{aligned}
J'_{\text{LS}}[n] &= \sum^{n}_{i=0}\lambda^{n-i}\left|d_{\text{I}}[i] - \ve{x}^T[i] \ve{w}_{\text{I}}[i-1] \ve{x}^H[i] \ve{w}_{\text{I}}^*[n] \ast h_\text{s}[i]\right|^2 \\
&= \sum^{n}_{i=0}\lambda^{n-i}\left|d_{\text{I}}[i] - \ve{z}^T[i] \ve{w}_{\text{I}}^*[n] \ast h_\text{s}[i]\right|^2 \\
&=\sum^{n}_{i=0}\lambda^{n-i}\left|e_{\text{I}}[i] \right|^2
\end{aligned}
\end{equation}
and introducing the new input vector \mbox{$\ve{z}[i]=\ve{x}^T[i] \ve{w}_{\text{I}}[i-1] \ve{x}^*[i]$}, we can overcome this limitation. Following the traditional \ac{RLS} derivation \cite{sayed2003fundamentals}, the IM2RLS algorithm to estimate the I-path \ac{IMD2} interference in the digital \ac{BB} becomes \eqref{eq:I_patch_IMD2_replica_final_equation_2}-\eqref{eq:coeff_update_2_final}: 
\begin{equation}\label{eq:I_patch_IMD2_replica_final_equation_2}
\begin{aligned}
\hat{y}_{\text{I}}[n] = \ve{z}^T[n] \ve{w}_{\text{I}}^*[n-1]  \ast h_\text{s}[n]
\end{aligned}
\end{equation}
\begin{equation}\label{eq:error}
\begin{aligned}
e_{\text{I}}[n] = d_{\text{I}}[n] - \hat{y}_{\text{I}}[n]
\end{aligned}
\end{equation}

\begin{equation}\label{eq:kalman_gain_final}
\begin{aligned}
\ve{k}[n]=\frac{\bm{P}[n-1] \ve{z}_{\text{f}}[n] }{\lambda + \ve{z}_{\text{f}}^H[n] \bm{P}[n-1] \ve{z}_{\text{f}}[n]}
\end{aligned}
\end{equation} 
\begin{equation}\label{eq:inv_autocorr_matrix_final}
\begin{aligned}
\bm{P}[n]=\frac{1}{\lambda}\left[\bm{P}[n-1] -\ve{k}[n] \ve{z}_{\text{f}}^H[n]  \bm{P}[n-1]\right]
\end{aligned}
\end{equation}
\begin{equation}\label{eq:coeff_update_2_final}
\begin{aligned}
\ve{w}_{\text{I}}[n]&=\ve{w}_{\text{I}}[n-1]+e_{\text{I}}[n] \ve{k}[n]
\end{aligned}
\end{equation} 
To avoid the channel-select filtering of each element in the vector \mbox{$\ve{z}_{\text{f}}[n] = \ve{z}[n] \ast h_s[n]$} which is mainly necessary to align the signals due to the CSF group delay, we introduce the signals \mbox{$x_{\text{f}}[n] = x[n] \ast h_s[n]$} and \mbox{$y_{\text{I}}'[n] = \ve{x}^T[n] \ve{w}_{\text{I}}[n-1]$}. Using the delay line vector \\ \mbox{$\ve{x}_{\text{f}}[n]=\left[x_{\text{f}}[n],x_{\text{f}}[n-1],\ldots,x_{\text{f}}[n-M+1]\right]^T$}, the vector \\ $\ve{z}_{\text{f}}[n]$ may be approximated by \mbox{$\ve{z}_{\text{f}}[n]\approx \left(y_{\text{I}}'[n]\ast h_s[n]\right)\ve{x}_{\text{f}}^*[n]$}. With this formulation, a fractional and non-constant group delay of the \ac{CSF} may be incorporated. In case if the group delay $\tau_{\text{g}}$ is constant, and an integer multiple of the sampling time (as e.g. in linear phase \ac{FIR} filters), the \ac{CSF} may be approximated by delaying the signal by \mbox{$\ve{z}_{\text{f}}[n] \approx \ve{x}^T[n-\tau_{\text{g}}] \ve{w}_{\text{I}}[n-1-\tau_{\text{g}}] \ve{x}^*[n-\tau_{\text{g}}]$}. In both approximations, the band-limiting effect of the \ac{CSF} on $\ve{z}_{\text{f}}[n]$ is ignored. However, this may be tolerated because due to the envelope-squaring operation in \eqref{eq:I_patch_IMD2_replica} which doubles the signal bandwidth, anyhow an \ac{OSF} of 2 is mandatory to avoid aliasing. 
Due to the fact, that the I-, and \mbox{Q-path} IMD2 interference differ only by a real valued scaling factor $\epsilon$ as derived in \eqref{eq:y_Tot_after_CSF_adaptive_filter}, the estimated I-path \ac{IMD2} replica may be used as a reference to estimate the \mbox{Q-path} IMD2 replica. This may be done by a linear \mbox{1-tap} RLS algorithm which uses the estimated \mbox{I-path} replica as reference input signal to estimate the \mbox{Q-path} IMD2 replica. In this case, the \mbox{1-tap} RLS estimates also a possible sign difference between the I-, and \mbox{Q-path} IMD2 interference. Consequently, only the sign of $\alpha_2^{\text{I}}$ has to be detected during calibration of the receiver which may be done by correlation. 
The replica signal generation \eqref{eq:I_patch_IMD2_replica_final_equation_2} is channel-select filtered which reduces the bandwidth of the replica signal to the bandwidth of the received \ac{LTE} signal.
\subsection{Second-Order Condition}
The complex Hessian \cite{Van_den_Bos_1,Schreier_2010} of the cost function \eqref{eq:LS_cost_function_sum} at the coefficient value \mbox{$\ve{w}_{\text{I}}=\ve{0}$} becomes 
\begin{equation}\label{eq:Hessian_matrix_with_DC}
\begin{aligned}
H_{\text{I}}&=\frac{\partial}{\partial \ve{w}_{\text{I}}}\left[\frac{\partial J_{\text{LS}}}{\partial \ve{w}_{\text{I}}^*}\right]^T|_{\ve{w}_{\text{I}}=\ve{0}} \\
&= \sum^{n}_{i=0}\lambda^{n-i} \left[-2\, d_{\text{I}}[i] \vx^*[i] \vx^T[i] \ast h_s[i]\right].
\end{aligned}
\end{equation}
If the desired signal $d_{\text{I}}[n]$ contains the DC (when the receiver has no DC filtering), then \mbox{$E\left\{d_{\text{I}}[n]\right\} \geq 0$} and the Hessian matrix becomes negative semi-definite like depicted with the local maximum in Fig. \ref{fig:LS_cost_function_3d_with_DC}. The usual choice of the zero-vector as initialization of $\ve{w}_{\text{I}}[-1]$ results in a zero-gain vector $\bm{k}[n]$ for all $n$. This is reasoned in the cost function \eqref{eq:LS_cost_function_sum} depicted in Fig. \ref{fig:LS_cost_function_3d_with_DC} which has a local maximum at $\ve{w}_{\text{I}}=\bm{0}$ and therefore a vanishing gradient. Consequently, the algorithm is initialized with \mbox{$\ve{w}_{\text{I}}[-1] \neq \bm{0}$} and the parameters \mbox{$0<<\lambda\leq1$}, and \mbox{$\bm{P}[-1] = \nu \, \bm{I}$} with \mbox{$\nu > 0$}. 
\subsection{DC Cancellation}
\begin{figure}[ht]
\centering
\input{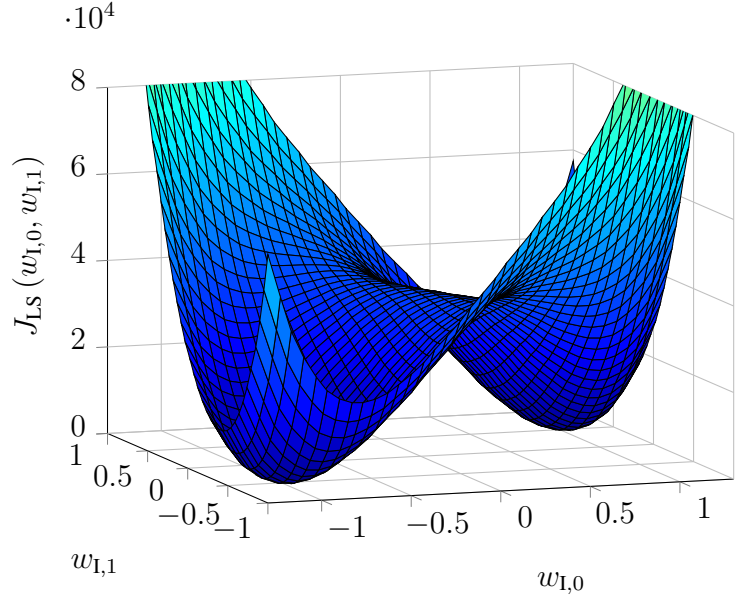}
\caption{Shape of the cost function \eqref{eq:LS_cost_function_sum} for white Gaussian input signals with $\lambda=1$ and for the two real valued coefficients \mbox{$\ve{h}=[1, 0.5]^T$}. The local maximum at $\ve{w}_{\text{I}}=\bm{0}$ (with DC) changed to a saddle-point because the DC filtering is applied.}
\label{fig:LS_cost_function_3d_without_DC}
\end{figure}

To employ an \ac{IMD2} interference replica without DC, the replica signal \eqref{eq:I_patch_IMD2_replica_final_equation_2} is filtered by the DC-notch filter \eqref{eq:DC_notch_filter_2}. The new error signal $e_{\text{AC,I}}[n] = d_{\text{AC,I}}[n] - \hat{y}_{\text{AC,I}}[n]$ with the DC-filtered signals is used in the update equation \eqref{eq:coeff_update_2_final_1}. Here, the introduced index AC indicates the DC filtered signals. The IM2RLS algorithm with DC-suppression can be summarized as \eqref{eq:I_patch_IMD2_replica_final_equation_3}-\eqref{eq:coeff_update_2_final_1}: 
\begin{equation}\label{eq:I_patch_IMD2_replica_final_equation_3}
\begin{aligned}
\hat{y}_{\text{I}}[n] = \ve{z}^T[n] \ve{w}_{\text{I}}^*[n-1]  \ast h_\text{s}[n]
\end{aligned}
\end{equation}
\begin{equation}\label{eq:DC_notch_filter_2}
\begin{aligned}
\hat{y}_{\text{AC,I}}[n] = a\, \hat{y}_{\text{AC,I}}[n-1] + \hat{y}_{\text{I}}[n] - \hat{y}_{\text{I}}[n-1].
\end{aligned}
\end{equation}
\begin{equation}\label{eq:error_1}
\begin{aligned}
e_{\text{AC,I}}[n] = d_{\text{AC,I}}[n] - \hat{y}_{\text{AC,I}}[n]
\end{aligned}
\end{equation}

\begin{equation}\label{eq:kalman_gain_final_1}
\begin{aligned}
\ve{k}[n]=\frac{\bm{P}[n-1] \ve{z}_{\text{f}}[n] }{\lambda + \ve{z}_{\text{f}}^H[n] \bm{P}[n-1] \ve{z}_{\text{f}}[n]}
\end{aligned}
\end{equation} 
\begin{equation}\label{eq:inv_autocorr_matrix_final_1}
\begin{aligned}
\bm{P}[n]=\frac{1}{\lambda}\left[\bm{P}[n-1] -\ve{k}[n] \ve{z}_{\text{f}}^H[n]  \bm{P}[n-1]\right]
\end{aligned}
\end{equation}
\begin{equation}\label{eq:coeff_update_2_final_1}
\begin{aligned}
\ve{w}_{\text{I}}[n]&=\ve{w}_{\text{I}}[n-1]+e_{\text{AC,I}}[n] \ve{k}[n]
\end{aligned}
\end{equation}
The parameter \mbox{$0<<a<1$} in \eqref{eq:DC_notch_filter_2} determines the sharpness of the DC-notch filter and is chosen as $a=0.998$. In case of DC filtering in the main receiver \mbox{$E\left\{d_{\text{I}}[n]\right\} = 0$}, and the Hessian matrix \eqref{eq:Hessian_matrix_with_DC} at $\ve{w}_{\text{I}} = \bm{0}$ is not positive semi-definite anymore. In this case, the local maximum becomes a saddle-point like depicted in Fig. \ref{fig:LS_cost_function_3d_without_DC}.  
Using $N_{\text{CSF}}$ as the number of coefficients of the \ac{CSF} impulse response, the computational complexity of the IM2RLS with DC-notch filter is \mbox{$13N_{\ve{w}}^2+5N_{\text{CSF}}+20N_{\ve{w}}+1$} real multiplications and $2 N_{\ve{w}}$ real divisions per iteration.

%
\subsection{Multiple Solutions of the IM2RLS Algorithm}
In the cost function shapes depicted in Fig.~\ref{fig:LS_cost_function_3d_with_DC} and Fig.~\ref{fig:LS_cost_function_3d_without_DC}, the estimated impulse response coefficients $w_{0}$ and $w_{1}$ (omitting the index I for the I-path) are constrained to be real valued. It can be observed that the two solutions \mbox{$\ve{w}_{0}=[1, 0.5]^T$}, and \mbox{$\ve{w}_{1} = [-1, -0.5]^T$} minimize the cost function. The existence of multiple solutions can be explained by the absolute-squaring nature of the \ac{IMD2} interference. 

\begin{figure}[ht]
\centering
\input{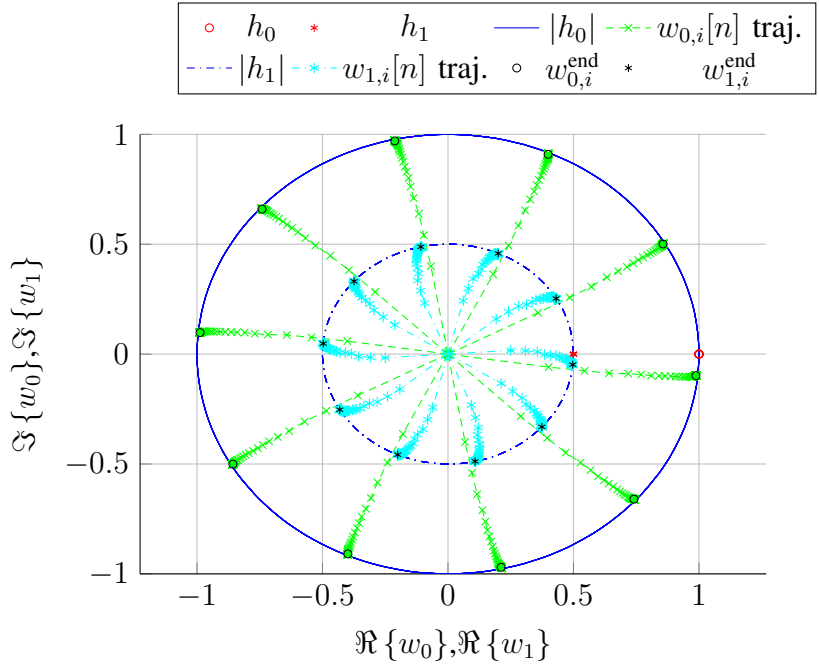}
\caption{Illustration of the initialization-dependent multiple solutions where the true coefficient values are \mbox{$\ve{h}=[1, 0.5]^T$}. The initial coefficient $w_0[-1]$ is initialized in a \mbox{10-point} grid around a circle with radius $1e-3$. The initial value of $h_1[-1]$ is always zero. With each initialization, the coefficients converge to the correct absolute value. All ten resulting estimated impulse response vectors $\ve{w}_i^{\text{end}}$ maintain the same group delay as $\ve{h}$.}
\label{fig:IM2RLS_multiple_solutions}
\end{figure}

If the coefficients are allowed to be complex valued, all coefficient pairs $\left\{w_{0}, w_{1}\right\}$ converge to $\left|w_{0}^{\text{end}}\right|=\left|h_{0}\right|$ and $\left|w_{1}^{\text{end}}\right|=\left|h_{1}\right|$. This scenario is visualized in Fig.~\ref{fig:IM2RLS_multiple_solutions} where the convergence of the coefficients with the ten different initializations \mbox{$\ve{w}_{i}[-1]=\left[1e-3,0\right]^T \text{exp}\left(j 2\pi/10 i\right)$} for \mbox{$i=0...9$} is depicted. Furthermore, each of the estimated coefficient vectors $\ve{w}_i^{\text{end}}=\left[w_{0,i}^{\text{end}},w_{1,i}^{\text{end}}\right]^T$ 
after convergence reach the group delay of the real system impulse response $\ve{h}$.   
%
%
%
%
%

%
\subsection{Performance of the IM2RLS with DC Suppression}
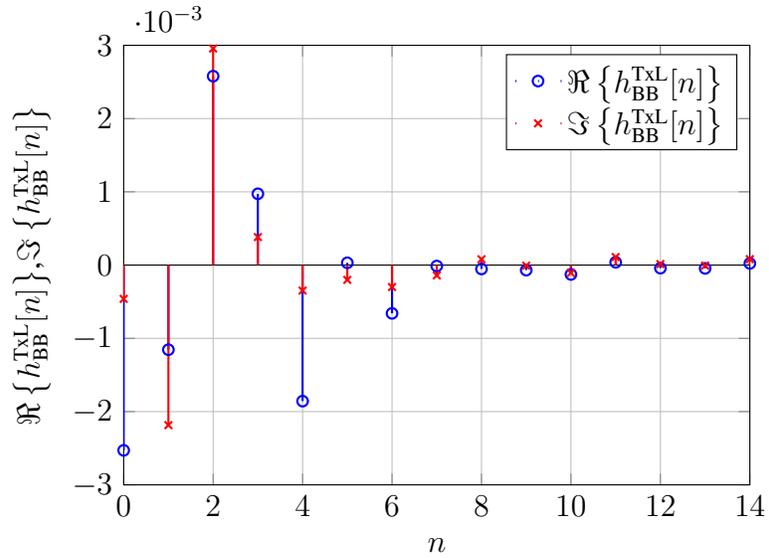
\begin{figure}[ht]
\centering
%
%
\begin{tikzpicture}

\begin{axis}[%
width=0.6*\columnwidth, 
height = .45\columnwidth,
xmin=0.000000,
xmax=14.000000,
ymin=-0.003000,
ymax=0.003000,
xlabel={$n$},
ylabel={$\Re\left\{h_{\text{BB}}^{\text{TxL}}[n]\right\}$,$\Im\left\{h_{\text{BB}}^{\text{TxL}}[n]\right\}$},
ytick={-3e-3,-2e-3,-1e-3,0,1e-3,2e-3,3e-3},
axis background/.style={fill=white},
xmajorgrids,
ymajorgrids,
legend style={legend cell align=left, align=left, draw=black}
]
\addplot[ycomb, color=blue, thick,  mark=o, mark options={solid, blue}] table[row sep=crcr] {%
0.000000	-0.002529\\
1.000000	-0.001155\\
2.000000	0.002579\\
3.000000	0.000973\\
4.000000	-0.001857\\
5.000000	0.000031\\
6.000000	-0.000659\\
7.000000	-0.000013\\
8.000000	-0.000053\\
9.000000	-0.000069\\
10.000000	-0.000128\\
11.000000	0.000038\\
12.000000	-0.000042\\
13.000000	-0.000045\\
14.000000	0.000023\\
};
\addlegendentry{$\Re\left\{h_{\text{BB}}^{\text{TxL}}[n]\right\}$}

\addplot [color=black, forget plot]
  table[row sep=crcr]{%
0.000000	0.000000\\
14.000000	0.000000\\
};
\addplot[ycomb, color=red, thick, mark=x, mark options={solid, red}] table[row sep=crcr] {%
0.000000	-0.000461\\
1.000000	-0.002185\\
2.000000	0.002956\\
3.000000	0.000382\\
4.000000	-0.000348\\
5.000000	-0.000202\\
6.000000	-0.000300\\
7.000000	-0.000141\\
8.000000	0.000080\\
9.000000	-0.000008\\
10.000000	-0.000107\\
11.000000	0.000110\\
12.000000	0.000014\\
13.000000	-0.000007\\
14.000000	0.000081\\
};
\addlegendentry{$\Im\left\{h_{\text{BB}}^{\text{TxL}}[n]\right\}$}

\end{axis}
\end{tikzpicture}%
\caption{Real and imaginary part of the 15-tap complex valued duplexer impulse response.}
\label{fig:dpx_h_coeff}
\end{figure}

In this section, the performance of the IM2RLS w/o and w/ DC cancellation is compared. 
In the first case, the receiver and the \ac{IMD2} replica generation of the IM2RLS do not use a DC cancellation. In this hypothetical example it is assumed that the \ac{IMD2} interference is the only DC source. In the second case, the receiver uses a DC suppression, and the IM2RLS the DC-notch filter. Both cases are compared within an \ac{FDD} scenario with full allocated \ac{LTE} signals using 10\,MHz bandwidth, QPSK modulation, short cyclic prefix, and an \ac{OSF} of 2. The frequency-selective duplexer stop-band impulse response shown in Fig. \ref{fig:dpx_h_coeff} is used in \eqref{eq:y_Tot_after_CSF} for the \ac{IMD2} interference generation. It is modeled with an \ac{FIR} system which has 15 complex valued coefficients (on the native LTE10 sampling rate of 15.36\,MHz) and a mean \ac{Tx}-to-\ac{Rx} isolation of 50\,dB \cite{Ericsson_1}. 
The resulting \ac{TxL} signal has a strong frequency-selectivity like indicated in Fig.~\ref{fig:spectrum}. The wanted \ac{Rx} signal power is at reference sensitivity level $P_{\text{Rx}}=-97\,\text{dBm}$ and the thermal noise floor is -104.5\,dBm within 10\,MHz bandwidth. The receiver \ac{NF} is 4.5\,dB which results in an receiver noise floor of -100\,dBm. The LNA gain is 20\,dB, and the two-tone mixer \ac{IIP2} is 50\,dBm. This results in an desensitization of the wanted \ac{Rx} signal from an \mbox{\ac{SNR}\,=\,3\,dB} to an \ac{SINR} of -1.4\,dB at $P_{\text{Tx}}\,=\,23\,\text{dBm}$. 
The I-path IMD2 interference is estimated by the IM2RLS using 15 taps, running at the sampling frequency of 30.72\,MHz \mbox{(OSF = 2)}. This means, the adaptive filter has less taps than the duplexer stop-band impulse response which has 30 complex valued coefficients at OSF = 2. The Q-path IMD2 replica is estimated by a linear 1-tap RLS (running at 30.72\,MHz sampling rate) which uses the I-path IMD2 replica as reference input. 

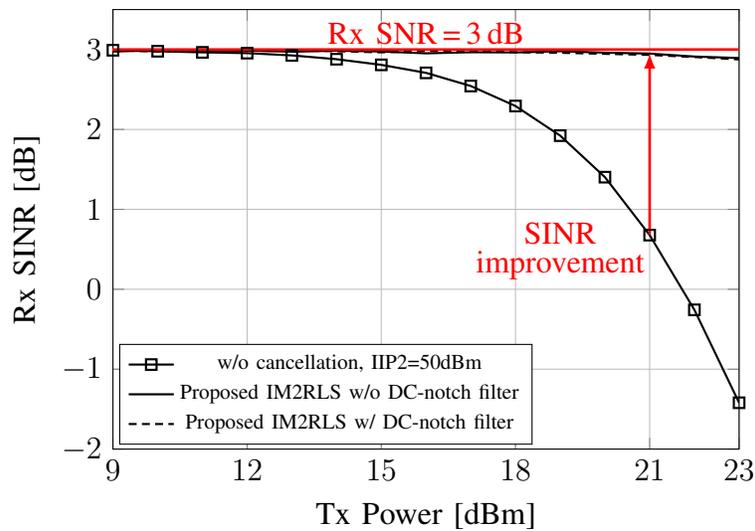
\begin{figure}[!ht]
\centering
\begin{tikzpicture}
\pgfmathsetlengthmacro{\textsizescale}{1}
\begin{axis}[ 
 width=0.6*\columnwidth, 
 height = .45\columnwidth,
 legend style={font=\scriptsize},
 xlabel= {Tx Power [dBm]}, 
 ylabel= {Rx SINR [dB]},
 xtick = {9,12,15,18,21,23},
 xmax = 23,
 xmin = 9,
 ymax = 3.5,
 ymin = -2,
 grid,
 legend style={at={(0.01,0.01)},anchor=south west}
]

%

%
%

\addplot[color=black, thick, mark = square] table[x index =0, y index =2] {plot_data/Tx_dBm_vs_Rx_SNIR_dB_Rx_-97dBm_IIP2_50dBm_no_DC_no_reg.dat};
\addlegendentry{w/o cancellation, IIP2=50dBm}
\addplot[color=black, thick] table[x index =0, y index =3] {plot_data/Tx_dBm_vs_Rx_SNIR_dB_Rx_-97dBm_IIP2_50dBm_with_DC_no_reg.dat};
\addlegendentry{Proposed IM2RLS w/o DC-notch filter}

\addplot[color=black, thick, style = densely dashed] table[x index =0, y index =3] {plot_data/Tx_dBm_vs_Rx_SNIR_dB_Rx_-97dBm_IIP2_50dBm_no_DC_no_reg.dat};
\addlegendentry{Proposed IM2RLS w/ DC-notch filter}

\addplot[color=red, line width=1] coordinates{(0,3) (23,3)};

\draw[line width=1, color = red, ->] (21,0.684) -- (21,2.954);
\node[scale=\textsizescale, color = red] at (19,0.7) {SINR};
\node[scale=\textsizescale, color = red] at (19,0.3) {improvement};
\node[scale=\textsizescale, color = red] at (16,3.2) {Rx SNR\,=\,3\,dB};
\end{axis} 

\end{tikzpicture}
\caption{Improvement of the Rx \ac{SINR} with the proposed \ac{IMD2} cancellation algorithms w/o and w/ using the DC-notch filter at different transmitter power levels. The mixer \ac{IIP2} is 50\,dBm and the wanted signal at the antenna has a power of $P_{\text{Rx}}$\,=\,-97\,dBm and a SNR of 3\,dB.}
\label{fig:SINR_algo_comparison_DC}
\end{figure}

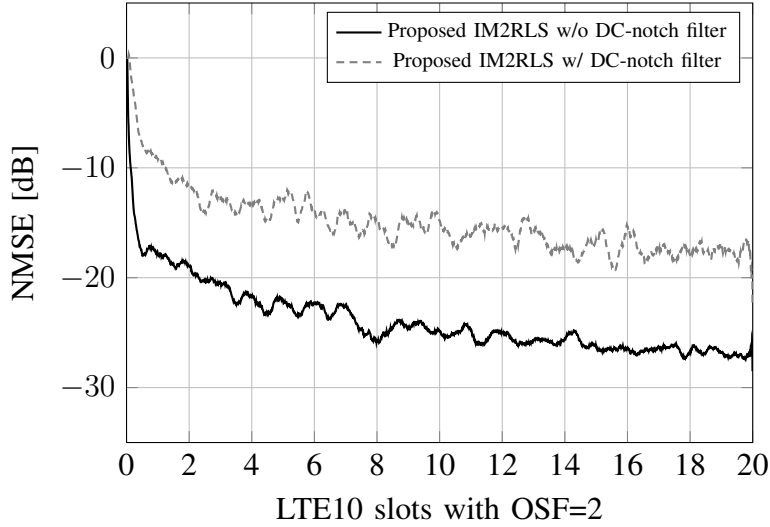
\begin{figure}[!ht]
\centering
\begin{tikzpicture}
\begin{axis}[
 width=0.6*\columnwidth, 
 height = .45\columnwidth,
 legend style={font=\scriptsize}, 
 xlabel= {LTE10 slots with OSF=2}, 
 ylabel= {NMSE [dB]},
 xmin = 0,
 xmax = 307200,
 ymin = -35,
 ymax = 5,
 xtick={0,30720,61440,92160,122880,153600,184320,215040,245760,276480,307200},  xticklabels={0,2,4,6,8,10,12,14,16,18,20},
 scaled x ticks = false,
 grid,
] 

%

\addplot[color=black, thick] table[x index =0, y index =1] {plot_data/cancellation_performance_smoothed_MA500_IIP2_50dBm_with_DC.dat};
\addlegendentry{Proposed IM2RLS w/o DC-notch filter}

\addplot[color=gray, thick, style = densely dashed] table[x index =0, y index =1] {plot_data/cancellation_performance_smoothed_MA500_IIP2_50dBm_without_DC.dat};
\addlegendentry{Proposed IM2RLS w/ DC-notch filter}

\end{axis} 
\end{tikzpicture}
\caption{Convergence of the IM2RLS w/o and w/ DC-notch filter for an \ac{LTE} transmit signal with 10\,MHz bandwidth, \ac{OSF} of 2 and $P_{\text{Tx}}$\,=\,23\,dBm. The wanted \ac{Rx} signal power at the antenna input is $P_{\text{Rx}}$\,=\,-97\,dBm and the \ac{Rx} \ac{SNR}\,=\,3\,dB. The mixer \ac{IIP2} is +50\,dBm which corresponds to an \ac{Rx} SNR desense of 4.4\,dB.}
\label{fig:convergence_algo_comparison_DC}
\end{figure}

The IM2RLS algorithm uses the forgetting-factor \mbox{$\lambda=0.9999$} and \mbox{$\bm{P}[-1]=100 \bm{I}$} as suggested in \cite{Isermann_1}. The \mbox{1-tap} RLS in the \mbox{Q-path} uses the same forgetting factor and the initial coefficient $p[-1] = 1e7$. The coefficient vector of the \mbox{I-path} IM2RLS algorithm is initialized with \mbox{$\ve{w}_{\text{I}}[-1]=[1e-6,0,0,...,0]^T$}, and the \mbox{1-tap} RLS with zero. Fig.~\ref{fig:SINR_algo_comparison_DC}. shows the steady state \ac{SINR} improvement at different transmit power levels for an \ac{IIP2} of +50\,dBm. It can be observed, that in both cases (w/o and w/ DC cancellation) the \ac{SINR} is improved nearly up to the \ac{Rx} \ac{SNR} of 3\,dB.
The convergence behavior at the transmit power of 23\,dBm is depicted in Fig.~\ref{fig:convergence_algo_comparison_DC}. For the hypothetical case that the receiver and the IM2RLS are using no DC suppression, the IM2RLS converges faster than with DC suppression. This is reasoned in the additional DC-\ac{IMD2} power which supports the algorithm to converge faster.
\begin{table}[ht]
\centering
		\caption{IIP2 improvement by digital cancellation}
    \begin{tabular}{ | c | c |c|c|}
    \hline
    IM2RLS Algorithm           &	$P_{\text{IMD2}}^{\text{CSF}}$ before	&	$P_{\text{IMD2}}^{\text{CSF}}$ after  & IIP2 after canc. \\ \hline \hline
    w/o DC cancellation		  	&	-77.5\,dBm	& -95.8\,dBm  &   68.4\,dBm  \\ \hline											
    w/ DC cancellation 			&	-77.5\,dBm  & -94.5\,dBm  &   67\,dBm \\ \hline
    \end{tabular}
	\label{table:IIP2_improvement_with_and_without_DC_no_reg}
\end{table}
The $\text{IIP2}$ improvement by the digital cancellation is summarized in Table \ref{table:IIP2_improvement_with_and_without_DC_no_reg} and may be calculated for the IM2RLS with DC-notch filter via
\begin{equation}\label{eq:IIP2_improvment_with_and_without_DC}
\begin{aligned}
\text{IIP2}_{\text{after canc.}}&= 2 P_{\text{RF}}^{\text{TxL}} - P_{\text{IM2, after canc.}}^{\text{CSF,LTE}} - 13.4\,\text{dB} \\
&= 2\cdot \left(23\,\text{dBm} - 50\,\text{dB} + 20\,\text{dB}\right)\\
& + 94.5\,\text{dBm} - 13.4\,\text{dB} = 67\,\text{dBm}.
\end{aligned}
\end{equation}
The $\text{IIP2}$ is improved from +50\,dBm to 68.4\,dBm and 67\,dBm by the digital cancellation with the IM2RLS w/o and w/ DC suppression, respectively. The correction factor of 13.4\,dB corrects the IMD2 power calculated with the \mbox{2-tone} formula, to the channel-select, and DC-filtered in-band \ac{IMD2} power for the LTE10 full allocation case \cite{Gebhard2017}. For the calculation of the \text{IIP2} improvement, the \ac{IMD2} power without DC is used in both cases. The derived IM2RLS algorithm with included DC-notch filter shows an excellent cancellation performance for a full allocated LTE10 transmit signal. However, for small bandwidth allocations like e.g. used in multi-cluster transmissions, the \ac{RLS}-type algorithm suffers from numerical instability due to the badly-conditioned autocorrelation matrix $\tilde{\bm{R}}$. To overcome this limitation, the regularized IM2RLS (\mbox{R-IM2RLS}) is derived in the next section. 
%

%

%
%
%
%
\section{Tikhonov Regularization of the nonlinear \ac{RLS}}\label{sec:R-IM2RLS}
To reduce the spectral \ac{OOB} emission of the \ac{LTE} signals, not all available subcarriers are allocated. A portion of the subcarriers at the band-edges (guard-band) are forced to zero which introduces correlation in the transmit \ac{BB} samples. E.g. in a 10\,MHz LTE signal a maximum of 600 out of 1024 subcarriers may be occupied by data \cite{LTEPublicSafety}. This correlation in the \ac{Tx} \ac{BB} signal $x_{\text{BB}}[n]$ leads to an badly-conditioned autocorrelation matrix $\bm{R}=E\left\{\vx_{\text{BB}}[n] \vx^H_{\text{BB}}[n]\right\}$ and respectively $\tilde{\bm{R}}=E\left\{\ve{z}_{\text{f}}[n] \ve{z}_{\text{f}}^H[n]\right\}$. Algorithms which need the estimation of the autocorrelation matrix or its inverse $\bm{P} = \bm{R}^{-1}$ to estimate the system coefficients either iteratively or in batch-mode, are sensitive to the condition number of $\bm{R}$ and may suffer from numerical instability if $\bm{R}$ is badly-conditioned. Because of this reason, a regularized version of the IM2RLS algorithm (R-IM2RLS) is derived in this section. 

A common method to overcome the problem of badly-conditioned autocorrelation matrices is regularization \cite{sayed2003fundamentals}. Adding a positive definite matrix to the estimated auto-correlation matrix in each iteration of the \ac{RLS} algorithm guarantees that the regularized autocorrelation matrix $\tilde{\bm{R}}'$ stays positive definite and maintains therefore the necessary condition for convergence and existence of \mbox{$\bm{P}=\tilde{\bm{R}}'^{-1}$} \cite{Gunnarsson_1}. 

This method is commonly known as Tikhonov-regularization where a matrix $\bm{L}$ is used for the regularization~\cite{huckle2012data}. 
%
%
%
%
By including a regularization term in the cost function \eqref{eq:LS_cost_function_sum_new}, the new cost function
\begin{equation}\label{eq:regularized_cost}
\begin{aligned}
J'_{\text{R}}[n] &=  \sum^{n}_{i=0}\lambda^{n-i}\left[\left|e_{\text{I}}[i] \right|^2 + \sigma \left\|\bm{L} \ve{w}_{\text{I}}[n]\right\|_2^2\right] \\
&=\sum^{n}_{i=0}\lambda^{n-i}\left[\left|e_{\text{I}}[i] \right|^2 + \sigma \, \ve{w}_{\text{I}}^T[n] \bm{L}^T \bm{L} \ve{w}_{\text{I}}^*[n]\right]
\end{aligned}
\end{equation}
is defined where $e_{\text{I}}[i]=d_{\text{I}}[i] - \ve{z}^T[i] \ve{w}_{\text{I}}^*[n] \ast h_\text{s}[i]$. The regularization parameter $\sigma \geq 0$ is used to adjust the regularization amount and the real valued matrix $\bm{L}$ is typically chosen as \mbox{$\bm{L}=\bm{I}$} (standard Tikhonov regularization), \mbox{$\bm{L}=\text{upperbidiag}\left(1,-1\right)$} (first order derivative), or 
\begin{equation}\label{eq:smoothing_matrix_L}
\bm{L}=
\begin{bmatrix}
-2 & 1 \\
 1 &-2 & 1 \\
  & 1 &-2 & 1     &         \\
   &   & \ddots & \ddots & \ddots \\
	& & & & \\
  &  &  &    1   & -2   
\end{bmatrix}
\end{equation}
(second order derivative) \cite{huckle2012data}. 
Using the Wirtinger calculus \cite{Brandwood_1} to obtain the gradient of the cost function \eqref{eq:regularized_cost}, and setting the gradient to zero results in
\begin{equation}\label{eq:regularized_wiener_equation}
\begin{aligned}
\underbrace{\left[\sum^{n}_{i=0}\lambda^{n-i} \left(\ve{z}_{\text{f}}[i] \ve{z}_{\text{f}}^H[i] + \sigma \bm{L}^T \bm{L} \right) \right]}_{\tilde{\bm{R}}'[n]} \ve{w}_{\text{I}}[n] = \underbrace{\sum^{n}_{i=0}\lambda^{n-i} d_{\text{I}}[i] \ve{z}_{\text{f}}[i]}_{\tilde{\ve{r}}[n]}.
\end{aligned}
\end{equation}
Reformulating the above equation leads to \mbox{$\ve{w}_{\text{I}}[n]=\tilde{\bm{R}}'^{-1}[n] \tilde{\ve{r}}[n]=\bm{P}[n] \tilde{\ve{r}}[n]$} which is solved recursively using the \ac{RLS} algorithm. By expressing the cross-correlation vector $\tilde{\ve{r}}[n]$ by its previous estimate $\tilde{\ve{r}}[n-1]$, a recursive estimation of the form 
\begin{equation}\label{eq:cross_corr_vec_regularized}
\begin{aligned}
\tilde{\ve{r}}[n] = \lambda \tilde{\ve{r}}[n-1] + d_{\text{I}}[n] \ve{z}_{\text{f}}[n]
\end{aligned}
\end{equation}
may be formulated. Similarly, a recursive estimation of the regularized autocorrelation matrix is obtained by
\begin{equation}\label{eq:iterative_update_autocorr_matrix_reg}
\begin{aligned}
\tilde{\bm{R}}'[n]&=\lambda \sum^{n-1}_{i=0}\lambda^{n-i-1} \left(\ve{z}_{\text{f}}[i] \ve{z}_{\text{f}}^H[i] + \sigma \bm{L}^T \bm{L} \right) \\
&+ \ve{z}_{\text{f}}[n] \ve{z}_{\text{f}}^H[n] + \sigma \bm{L}^T \bm{L} \\
&= \lambda \tilde{\bm{R}}'[n-1] + \sigma \bm{L}^T \bm{L} + \ve{z}_{\text{f}}[n] \ve{z}_{\text{f}}^H[n].
\end{aligned}
\end{equation} 
Substituting $\bm{\Omega}[n]^{-1}=\lambda \tilde{\bm{R}}'[n-1] + \sigma \bm{L}^T \bm{L}$ into \eqref{eq:iterative_update_autocorr_matrix_reg}, the matrix $\bm{P}[n]=\tilde{\bm{R}}'^{-1}[n]$ becomes 
\begin{equation}\label{eq:inv_autocorr_matrix_reg}
\begin{aligned}
\bm{P}[n]= \left[\bm{\Omega}[n]^{-1} + \ve{z}_{\text{f}}[n] \ve{z}_{\text{f}}^H[n]\right]^{-1}.
\end{aligned}
\end{equation} 
After applying the matrix inversion lemma 
\begin{equation}\label{eq:matrix_inversion_lemma}
\begin{aligned}
\left(\bm{A}+\bm{B}\bm{C}\bm{D}\right)^{-1}=\bm{A}^{-1}-\bm{A}^{-1}\bm{B}\left(\bm{C}^{-1}+\bm{D} \bm{A}^{-1} \bm{B}\right)^{-1} \bm{D} \bm{A}^{-1}
\end{aligned}
\end{equation} 
to avoid the matrix inversion, \eqref{eq:inv_autocorr_matrix_reg} may be formulated as
\begin{equation}\label{eq:inv_autocorr_matrix_reg2}
\begin{aligned}
\bm{P}[n]= \bm{\Omega}[n] - \bm{k}[n] \ve{z}_{\text{f}}^H[n] \bm{\Omega}[n]
\end{aligned}
\end{equation} 
using the gain vector
\begin{equation}\label{eq:kalman_gain_reg}
\begin{aligned}
\bm{k}[n]=\frac{ \bm{\Omega}[n] \ve{z}_{\text{f}}[n]}{ 1+ \ve{z}_{\text{f}}^H[n] \bm{\Omega}[n]  \ve{z}_{\text{f}}[n] }.
\end{aligned}
\end{equation}
For the inversion 
\begin{equation}\label{eq:omega}
\begin{aligned}
\bm{\Omega}[n]=\left[\lambda \bm{P}^{-1}[n-1] + \sigma \bm{L}^T \bm{L}\right]^{-1},
\end{aligned}
\end{equation} 
again the matrix inversion lemma is applyied which yields
\begin{equation}\label{eq:omega2}
\begin{aligned}
\bm{\Omega}[n]=\frac{1}{\lambda}\left(\bm{P}[n-1] - \bm{\Sigma}[n] \bm{L} \bm{P}[n-1]  \right)
\end{aligned}
\end{equation} 
where the substitution
\begin{equation}\label{eq:S_matrix}
\begin{aligned}
\bm{\Sigma}[n]= \sigma \bm{P}[n-1] \bm{L}^T \left[ \lambda \bm{I} + \sigma \bm{L} \bm{P}[n-1] \bm{L}^T \right]^{-1}
\end{aligned}
\end{equation} 
is used. After rearranging \eqref{eq:S_matrix}, the expression 
\begin{equation}\label{eq:S_matrix_2}
\begin{aligned}
\bm{\Sigma}[n]&= \frac{\sigma}{\lambda} \left(\bm{P}[n-1] - \bm{\Sigma}[n] \bm{L} \bm{P}[n-1]  \right) \bm{L}^T \\
&= \sigma \bm{\Omega}[n] \bm{L}^T
\end{aligned}
\end{equation} 
is obtained. Unfortunately, the calculation of $\bm{\Sigma}[n]$ in \eqref{eq:S_matrix} and therefore $\bm{\Omega}[n]$ still includes a matrix inversion after applying the matrix inversion lemma. However, by decomposing the matrix $\bm{L}^T \bm{L}$ in \eqref{eq:omega} into a sum of $V$ dyads \cite{Dokoupil_1} 
\begin{equation}\label{eq:omega_dyads}
\begin{aligned}
\bm{\Omega}[n]=\left[\lambda \bm{P}^{-1}[n-1] + \sigma \sum^{V}_{k=1} \ve{p}_{k,1} \ve{p}_{k,2}^T \right]^{-1},
\end{aligned}
\end{equation} 
applying the matrix inversion lemma results in the recursive calculation of \eqref{eq:omega_dyads} via
\begin{equation}\label{eq:omega_dyads_recursive}
\begin{aligned}
\bm{\Omega}_{k}[n]=\bm{\Omega}_{k-1}[n] - \frac{\bm{\Omega}_{k-1}[n] \ve{p}_{k,1}}{\frac{1}{\sigma}+\ve{p}_{k,2}^T \bm{\Omega}_{k-1}[n] \ve{p}_{k,1}} \ve{p}_{k,2}^T \bm{\Omega}_{k-1}[n]
\end{aligned}
\end{equation} 
for $k=1 \ldots V$ in each iteration $n$ and $\bm{\Omega}_{0}[n]=\frac{1}{\lambda} \bm{P}[n-1]$. Reformulating \eqref{eq:kalman_gain_reg} yields 
\begin{equation}\label{eq:kalman_gain_reformulated}
\begin{aligned}
\ve{k}[n] = \bm{P}[n] \ve{z}_{\text{f}}[n].
\end{aligned}
\end{equation}
The recursive update of the coefficient vector $\ve{w}_{\text{I}}[n]$ is obtained by inserting \eqref{eq:inv_autocorr_matrix_reg2}, \eqref{eq:cross_corr_vec_regularized}, \eqref{eq:kalman_gain_reformulated}, \eqref{eq:omega2}  and \eqref{eq:S_matrix_2} into \mbox{$\ve{w}_{\text{I}}[n] = \bm{P}[n] \tilde{\ve{r}}[n]$}. The final nonlinear R-IM2RLS algorithm to estimate the I-path \ac{IMD2} interference is summarized by \eqref{eq:I_patch_IMD2_replica_final_equation_4}-\eqref{eq:coeff_update_2_final_2}:
\begin{equation}\label{eq:I_patch_IMD2_replica_final_equation_4}
\begin{aligned}
\hat{y}_{\text{I}}[n] = \ve{z}^T[n] \ve{w}_{\text{I}}^*[n-1]  \ast h_\text{s}[n]
\end{aligned}
\end{equation}
%
%
\begin{equation}\label{eq:error_2}
\begin{aligned}
e_{\text{I}}[n] = d_{\text{I}}[n] - \hat{y}_{\text{I}}[n]
\end{aligned}
\end{equation}
\begin{equation}\label{eq:omega_dyads_recursive_1}
\begin{aligned}
\bm{\Omega}_{k}[n]=\bm{\Omega}_{k-1}[n] - \frac{\bm{\Omega}_{k-1}[n] \ve{p}_{k,1}}{\frac{1}{\sigma}+\ve{p}_{k,2}^T \bm{\Omega}_{k-1}[n] \ve{p}_{k,1}} \ve{p}_{k,2}^T \bm{\Omega}_{k-1}[n]
\end{aligned}
\end{equation} 
\begin{equation}\label{eq:kalman_gain_reg_1}
\begin{aligned}
\bm{k}[n]=\frac{ \bm{\Omega}_{V}[n] \ve{z}_{\text{f}}[n]}{ 1+ \ve{z}_{\text{f}}^H[n] \bm{\Omega}_{V}[n]  \ve{z}_{\text{f}}[n] }.
\end{aligned}
\end{equation}
\begin{equation}\label{eq:inv_autocorr_matrix_reg2_1}
\begin{aligned}
\bm{P}[n]= \bm{\Omega}_{V}[n] - \bm{k}[n] \ve{z}_{\text{f}}^H[n] \bm{\Omega}_{V}[n]
\end{aligned}
\end{equation}
\begin{equation}\label{eq:S_matrix_2_1}
\begin{aligned}
\bm{\Sigma}[n]= \sigma \bm{\Omega}_{V}[n] \bm{L}^T
\end{aligned}
\end{equation}
\begin{equation}\label{eq:coeff_update_2_final_2}
\begin{aligned}
\ve{w}_{\text{I}}[n]&= \left[\bm{I} - \left(\bm{I}-\ve{k}[n] \ve{z}_{\text{f}}^H[n]\right)\bm{\Sigma}[n] \bm{L}\right] \ve{w}_{\text{I}}[n-1] + \ve{k}[n] e_{\text{I}}[n]
\end{aligned}
\end{equation}
The proposed algorithm is initialized with $\ve{w}_{\text{I}}[-1] \neq \bm{0}$, \mbox{$0<<\lambda\leq1$} and $\bm{P}[-1] = \nu \, \bm{I}$ with $\nu > 0$.  
When the DC suppression is used, then the R-IM2RLS update equations become \eqref{eq:I_patch_IMD2_replica_final_equation_5}-\eqref{eq:coeff_update_2_final_3}:
\begin{equation}\label{eq:I_patch_IMD2_replica_final_equation_5}
\begin{aligned}
\hat{y}_{\text{I}}[n] = \ve{z}^T[n] \ve{w}_{\text{I}}^*[n-1]  \ast h_\text{s}[n]
\end{aligned}
\end{equation}
\begin{equation}\label{eq:DC_notch_filter_3}
\begin{aligned}
\hat{y}_{\text{AC,I}}[n] = 0.998\, \hat{y}_{\text{AC,I}}[n-1] + \hat{y}_{\text{I}}[n] - \hat{y}_{\text{I}}[n-1]
\end{aligned}
\end{equation}
\begin{equation}\label{eq:error_regularization_with_DC_no_DC}
\begin{aligned}
e_{\text{AC,I}}[n] = d_{\text{AC,I}}[n] - \hat{y}_{\text{AC,I}}[n]
\end{aligned}
\end{equation}
\begin{equation}\label{eq:omega_dyads_recursive_2}
\begin{aligned}
\bm{\Omega}_{k}[n]=\bm{\Omega}_{k-1}[n] - \frac{\bm{\Omega}_{k-1}[n] \ve{p}_{k,1}}{\frac{1}{\sigma}+\ve{p}_{k,2}^T \bm{\Omega}_{k-1}[n] \ve{p}_{k,1}} \ve{p}_{k,2}^T \bm{\Omega}_{k-1}[n]
\end{aligned}
\end{equation} 
\begin{equation}\label{eq:kalman_gain_reg_2}
\begin{aligned}
\bm{k}[n]=\frac{ \bm{\Omega}_{V}[n] \ve{z}_{\text{f}}[n]}{ 1+ \ve{z}_{\text{f}}^H[n] \bm{\Omega}_{V}[n]  \ve{z}_{\text{f}}[n] }.
\end{aligned}
\end{equation}
\begin{equation}\label{eq:inv_autocorr_matrix_reg2_2}
\begin{aligned}
\bm{P}[n]= \bm{\Omega}_{V}[n] - \bm{k}[n] \ve{z}_{\text{f}}^H[n] \bm{\Omega}_{V}[n]
\end{aligned}
\end{equation}
\begin{equation}\label{eq:S_matrix_2_2}
\begin{aligned}
\bm{\Sigma}[n]= \sigma \bm{\Omega}_{V}[n] \bm{L}^T
\end{aligned}
\end{equation}
\begin{equation}\label{eq:coeff_update_2_final_3}
\begin{aligned}
\ve{w}_{\text{I}}[n]&= \left[\bm{I} - \left(\bm{I}-\ve{k}[n] \ve{z}_{\text{f}}^H[n]\right)\bm{\Sigma}[n] \bm{L}\right] \ve{w}_{\text{I}}[n-1] \\
&+ \ve{k}[n] e_{\text{AC,I}}[n]
\end{aligned}
\end{equation}
The DC-notch filter \eqref{eq:DC_notch_filter_3} is used to remove the DC from the \ac{IMD2} replica \eqref{eq:I_patch_IMD2_replica_final_equation_5}. The complexity of the R-IM2RLS with DC-notch filter and $L=\sigma \bm{I}$ is \mbox{$8 N_{\ve{w}}^3 + 21N_{\ve{w}}^2+5N_{\text{CSF}}+18N_{\ve{w}}+1$} real multiplications and $2 N_{\ve{w}}^2+2N_{\ve{w}}$ real divisions per iteration.
%
%

\section{Simulation environment}\label{sec:simulations}
The performance of the R-IM2RLS algorithm with the three above mentioned regularization matrices $\bm{L}$ is evaluated with an \ac{FDD} scenario using an LTE10 multi-cluster intra-band \ac{Tx} signal which has a native sampling frequency of $f_s=15.36\,\text{MHz}$, QPSK modulation and short cyclic prefix. The \ac{IMD2} interference in the \mbox{I-path} is estimated by the \mbox{R-IM2RLS}, while the \mbox{Q-path} \ac{IMD2} is estimated by a linear \mbox{1-tap} RLS which uses the \mbox{I-path} IMD2 replica as reference input. 
The resulting multi-cluster \ac{TxL} signal has a strong frequency-selectivity like indicated in Fig. \ref{fig:spectrum_clustered_Tx}. The R-IM2RLS in the \mbox{I-path} has 15 taps and runs on the higher sampling rate of 30.72\,MHz due to the \ac{OSF} of 2. This means, the adaptive filter has less taps than the impulse response which is estimated. The linear 1-tap \mbox{Q-path} RLS runs also on the sampling rate of 30.72\,MHz.   
The received signal $d[n]$ is DC filtered and the proposed algorithm is using the DC-notch filter to suppress the DC of the \ac{IMD2} replica signal. The wanted \ac{Rx} signal has a power of \mbox{$P_{\text{Rx}}$\,=\,-97\,dBm} at the antenna with an \ac{SNR} of 3\,dB. The assumed \ac{Rx} mixer \ac{IIP2} is +60\,dBm which corresponds to an \ac{Rx} SNR desense of 1\,dB for the specific intra-band multi-cluster transmit signal at 23\,dBm power level. The thermal noise floor of the receiver is assumed at -104.5\,dBm per 10\,MHz and the receiver \ac{NF} is 4.5\,dB. The resulting receiver noise floor and \ac{Rx} power with 20\,dB \ac{LNA} gain is at \mbox{-80\,dBm}\,$\widehat{=}$\,\mbox{-108.2\,dBm/15\,kHz} and -77\,dBm respectively. The spectrum of the signals at $P_{\text{Tx}}=23\,\text{dBm}$ is depicted in \mbox{Fig. \ref{fig:spectrum_clustered_Tx}}. It can be observed, that the resulting \ac{IMD2} interference $y_{\text{BB}}^{\text{IMD2}}$ is mostly below the receiver noise floor but still leads to an \ac{SNR} degradation of 1\,dB. The depicted interference replica is estimated by the R-IM2RLS with the regularization $\bm{L}=3e-7\,\bm{I}$.  
\begin{figure}[ht]
\centering
\begin{tikzpicture}
	\begin{axis}[
	    legend style={font=\scriptsize},
			width=0.6*\columnwidth, 
			height = .45\columnwidth,
			legend columns = {2},
			grid, 
			xlabel={$f$ [MHz]}, 
			ylabel={PSD [dBm/15\,kHz]},
			xmin = -8,
			xmax = 8,
			ymin = -200,
			ymax = -20,
			ytick={-200,-180,-160,-140,-120,...,-60,-40,-20},
			legend style={at={(0.295,0.01)},anchor=south west}
			]
			 
			
			
			\addplot[color=black, thick, mark = x, mark repeat={5}] table[x index =0, y index =3] {plot_data/spectrum_clustered_Tx_IIP_60dBm.dat};
			\addlegendentry{$y_{\text{BB}}^{\text{TxL}}$}
			
			
			\addplot[color=red, thick] table[x index =0, y index =1] {plot_data/spectrum_clustered_Tx_IIP_60dBm.dat};
			\addlegendentry{$y_{\text{BB}}^{\text{Tot}}$}
			
			\addplot[color=green, thick, style = dashed] table[x index =0, y index =2] {plot_data/spectrum_clustered_Tx_IIP_60dBm.dat};
			\addlegendentry{$y_{\text{BB}}^{\text{Rx}}$}
			
						\addplot[color=blue, thick] table[x index =0, y index =4] {plot_data/spectrum_clustered_Tx_IIP_60dBm.dat};
			\addlegendentry{$y_{\text{BB}}^{\text{IMD2}}$}
			
									\addplot[color=blue, thick, style = densely dashed, mark = x, mark repeat={5}] table[x index =0, y index =7] {plot_data/spectrum_clustered_Tx_IIP_60dBm.dat};
			\addlegendentry{$\hat{y}_{\text{BB}}^{\text{IMD2}}$}
			
			\addplot[color=black, thick, style = densely dotted] table[x index =0, y index =6] {plot_data/spectrum_clustered_Tx_IIP_60dBm.dat};
			\addlegendentry{Noise}
			
  \end{axis}
\end{tikzpicture}
\caption{Equivalent \ac{BB} spectrum of the frequency-selective \ac{Tx} leakage signal $y_{\text{BB}}^{\text{TxL}}$ (the corresponding passband signal is located at $f_{\text{Tx}}$) and the total received signal $y_{\text{BB}}^{\text{Tot}}$ after amplification with 20\,dB \ac{LNA} gain. The wanted Rx signal with SNR\,=\,3\,dB, and the receiver noise floor after amplification with 20\,dB \ac{LNA} gain are at -77\,dBm and \mbox{-80\,dBm}\,$\widehat{=}$\,\mbox{-108.2\,dBm/15\,kHz} respectively. The total received signal contains the DC-, and channel-select filtered \ac{IMD2} interference at $P_{\text{Tx}}$\,=\,23\,dBm and the IIP2 is 60\,dBm.
}
\label{fig:spectrum_clustered_Tx}
\end{figure}
The multi-cluster LTE10 \ac{Tx} signal uses 21/50 RBs (252 subcarriers from 1024), which means hat 3.78\,MHz of the available 9.015\,MHz are allocated. With an \ac{OSF} of 2 this corresponds to an allocated bandwidth-to-sampling-rate ratio of 3.78/30.72 = 0.12 which introduces a high correlation in the transmit \ac{BB} samples. The resulting condition number \mbox{$\text{cond}(\tilde{\bm{R}})$} of the \mbox{$15 \times 15$} dimensional autocorrelation matrix \mbox{$\tilde{\bm{R}}=E\left\{\ve{z}_{\text{f}}\ve{z}_{\text{f}}^H\right\}$} is in the order of \mbox{$10^7$} which results in a bad conditioned estimation, and may lead to numerical problems. The regularization of the R-IM2RLS improves numerical estimation of the matrix $\bm{P}[n]$ by lowering the condition number of the regularized matrix $\tilde{\bm{R}}'$. 
\subsection{IMD2 Self-Interference of a Multi-Cluster Tx Signal}
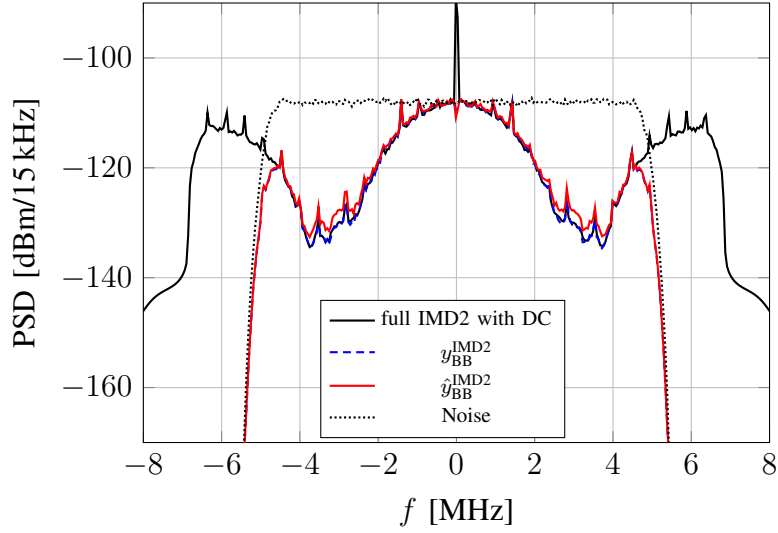
\begin{figure}[ht]
\centering
\begin{tikzpicture}
	\begin{axis}[
	    legend style={font=\scriptsize},
			width=0.6*\columnwidth, 
			height = .45\columnwidth,
			legend columns = {1},
			grid, 
			xlabel={$f$ [MHz]}, 
			ylabel={PSD [dBm/15\,kHz]},
			xmin = -8,
			xmax = 8,
			ymin = -170,
			ymax = -90,
			ytick={-200,-180,-160,-140,-120,...,-60,-40,-20},
			legend style={at={(0.2825,0.01)},anchor=south west}
			]
			 
			\addplot[color=black, thick] table[x index =0, y index =10] {plot_data/spectrum_IMD2_full_BW_IIP2_60dBm_clustered_Tx_reg_with_sigma_I.dat};
			\addlegendentry{full IMD2 with DC}
			
						\addplot[color=blue, thick, style = densely dashed] table[x index =0, y index =4] {plot_data/spectrum_IMD2_full_BW_IIP2_60dBm_clustered_Tx_reg_with_sigma_I.dat};
			\addlegendentry{$y_{\text{BB}}^{\text{IMD2}}$}
			
									\addplot[color=red, thick] table[x index =0, y index =7] {plot_data/spectrum_IMD2_full_BW_IIP2_60dBm_clustered_Tx_reg_with_sigma_I.dat};
			\addlegendentry{$\hat{y}_{\text{BB}}^{\text{IMD2}}$}
			
			\addplot[color=black, thick, style = densely dotted] table[x index =0, y index =6] {plot_data/spectrum_IMD2_full_BW_IIP2_60dBm_clustered_Tx_reg_with_sigma_I.dat};
			\addlegendentry{Noise}
			
  \end{axis}
\end{tikzpicture}
\caption{Generated \ac{IMD2} interference with the bandwidth of 13.68\,MHz at $P_{\text{Tx}}=23\,\text{dBm}$. The resulting in-band \ac{BB} \ac{IMD2} interference \mbox{$y_{\text{BB}}^{\text{IMD2}}$} after the \ac{CSF} and DC-removal is below the receiver noise floor. The R-IM2RLS estimates the \ac{IMD2} interference down to 20\,dB below the noise floor. 
}
\label{fig:spectrum_clustered_Tx_IMD2}
\end{figure}

For the estimation of the resulting \ac{IMD2} interference bandwidth, the bandwidth between the minimum and maximum allocated subcarrier in the multi-cluster \ac{Tx} signal is of interest. In the used clustered LTE10 transmit signal the allocated \acp{RB} are $\left\{9-11,29-46\right\}$ with a numbering from left to right and the total number of 50 \acp{RB}. For the \ac{IMD2} bandwidth estimation the resulting bandwidth between the lowest allocated subcarrier (\ac{RB} 9) and the upper edge (\ac{RB} 46) of the allocated \acp{RB} is \mbox{$\left(3+17+18\right)\cdot 12 \cdot 15\,\text{kHz} = 6.84\,\text{MHz}$}. Each \ac{RB} has 12 subcarriers and 15\,kHz subcarrier spacing. The resulting \ac{IMD2} interference bandwidth is $2\times 6.84\,\text{MHz}=13.68\, \text{MHz}$ which means that a small portion of the \ac{IMD2} interference is suppressed by the \ac{CSF}. The full \ac{IMD2} interference including the DC, the \ac{IMD2} interference after the \ac{CSF} and DC-removal, and the estimated \ac{IMD2} replica are visualized in Fig. \ref{fig:spectrum_clustered_Tx_IMD2}. It can be observed, that the \mbox{R-IM2RLS} is able to estimate the \ac{IMD2} interference down to 20\,dB below the receiver noise floor. 
\subsection{Numerical Simulation Results}
\begin{figure}[!ht]
\centering
\begin{tikzpicture}
\begin{axis}[
 width=0.6*\columnwidth, 
 height = .35\columnwidth,
 legend style={font=\scriptsize}, 
 xlabel= {LTE10 slots with OSF=2}, 
 ylabel= {Condition number of $\tilde{\bm{R}}'$},
 xmin = 0,
 xmax = 184320,
 ymin = 0,
 ymax = 1000,
 xtick={0,30720,61440,92160,122880,153600,184320},  
 xticklabels={0,2,4,6,8,10,12},
 scaled x ticks = false,
 grid]

\addplot[color=black, thick] table[x index =0, y index =2] {plot_data/Rxx_condition_number_over_time_IIP2_60dBm_no_DC_clustered_Tx_no_reg.dat};
\addlegendentry{IM2RLS w/o regularization}

\addplot[color=black, thick, style = densely dashed] table[x index =0, y index =2] {plot_data/Rxx_condition_number_over_time_IIP2_60_no_DC_clustered_Tx_reg_with_sigma_I.dat};
\addlegendentry{R-IM2RLS, $\bm{L}=3e-7\,\bm{I}$}

\addplot[color=black, thick, style = densely dotted] table[x index =0, y index =2] {plot_data/Rxx_condition_number_over_time_IIP2_60_no_DC_clustered_Tx_reg_with_sigma_1_-1.dat};
\addlegendentry{R-IM2RLS, $\bm{L}=3e-7\, \text{upperbidiag}\left(1,-1\right)$}

\addplot[color=gray, thick] table[x index =0, y index =2] {plot_data/Rxx_condition_number_over_time_IIP2_60_no_DC_clustered_Tx_smooth_matrix_L.dat};
\addlegendentry{R-IM2RLS using \eqref{eq:smoothing_matrix_L} and $\sigma = 3e-7$}

\end{axis} 
\end{tikzpicture}
\caption{Evolution of the condition number of $\tilde{\bm{R}}'[n] = \bm{P}^{-1}[n]$ for a clustered allocation like depicted in Fig.~\ref{fig:spectrum_clustered_Tx} and 23\,dBm transmit power. The condition number of $\tilde{\bm{R}}=E\left\{\ve{z} \ve{z}^H\right\}$ without regularization is in the order of $10^7$ to $10^8$.}
\label{fig:Rxx_cond_number}
\end{figure}
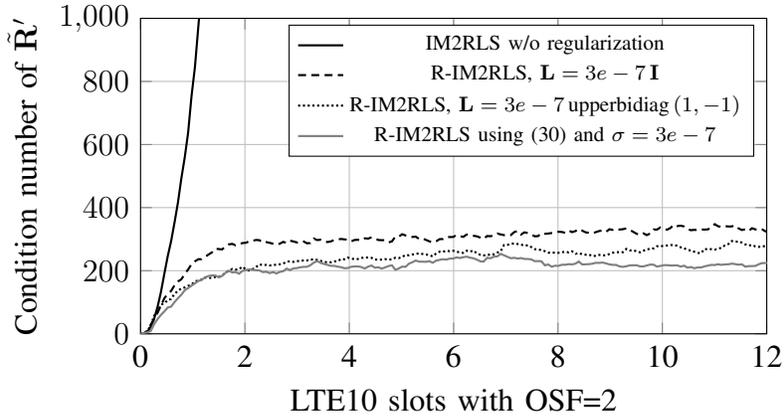

In the following simulation results, the \ac{IMD2} self-interference cancellation performance in case of an intra-band multi-cluster Tx signal, using the R-IM2RLS algorithm \mbox{\eqref{eq:I_patch_IMD2_replica_final_equation_5}-\eqref{eq:coeff_update_2_final_3}} using the DC-notch filter with different regularization matrices is evaluated. The forgetting factor of the \mbox{R-IM2RLS} is chosen as \mbox{$\lambda = 0.9999$},  \mbox{$\bm{P}[-1]=100 \bm{I}$}, and the regularization constant \mbox{$\sigma = 3e-7$}. The \mbox{1-tap} RLS in the \mbox{Q-path} uses the same forgetting factor but the initial coefficient $p[-1]=1e7$. The coefficient vector of the \mbox{R-IM2RLS} is initialized with \mbox{$\ve{w}_{\text{I}}[-1]=[1e-6,0,0,...,0]^T$} for the \mbox{I-path}, and the \mbox{1-tap} \mbox{Q-path} RLS is initialized with zero. The performance is evaluated for the different regularization matrices \mbox{$\bm{L}=3e-7\, \bm{I}$} (Tikhonov regularization),  \mbox{$\bm{L}=3e-7\, \text{upperbidiag}\left(1,-1\right)$} (first order derivative smoothing matrix), and \mbox{$\bm{L}=3e-7\, \text{diag}\left(1,-2,1\right)$} (second order derivative smoothing matrix). The IM2RLS without regularization is not included in the comparison due to numerical instability reasoned by the extremely high condition number of $\tilde{\bm{R}}$ which is in the order of $10^7$. The performance of the \mbox{R-IM2RLS} is compared with the recently published LMS-type algorithm (IM2LMS)~\cite{Gebhard2017}.  
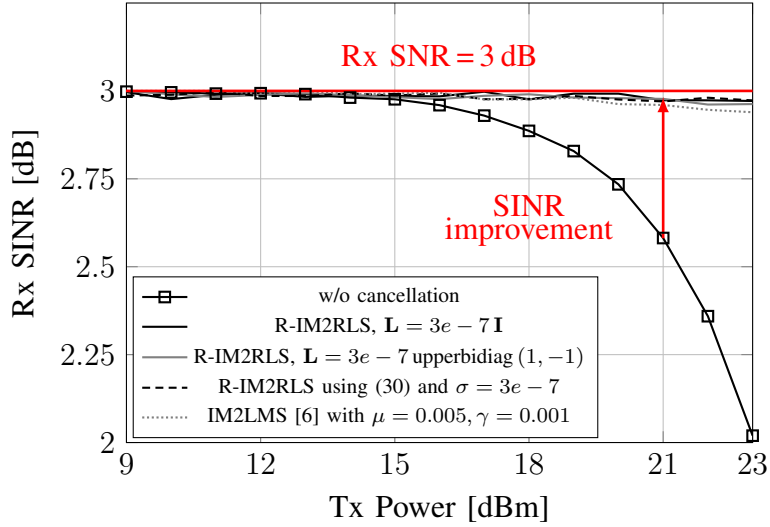
\begin{figure}[!ht]
\centering
\begin{tikzpicture}
\pgfmathsetlengthmacro{\textsizescale}{1}
\begin{axis}[ 
 width=0.6*\columnwidth, 
 height = .45\columnwidth,
 legend style={font=\scriptsize},
 xlabel= {Tx Power [dBm]}, 
 ylabel= {Rx SINR [dB]},
 xtick = {9,12,15,18,21,23},
 ytick={2,2.25,2.5,2.75,3},
 xmax = 23,
 xmin = 9,
 ymax = 3.25,
 ymin = 2,
 grid,
 legend style={at={(0.01,0.01)},anchor=south west}
]

%

\addplot[color=black, thick, mark = square] table[x index =0, y index =2] {plot_data/Tx_dBm_vs_Rx_SNIR_dB_Rx_-97dBm_IIP2_60dBm_no_DC_cluster_Tx_smooth_matrix_L.dat};
\addlegendentry{w/o cancellation}


\addplot[color=black, thick] table[x index =0, y index =3] {plot_data/Tx_dBm_vs_Rx_SNIR_dB_Rx_-97dBm_IIP2_60dBm_no_DC_cluster_Tx_reg_with_sigma_I_1_tap_for_Q_IMD2.dat};
\addlegendentry{R-IM2RLS, $\bm{L}=3e-7\,\bm{I}$}


\addplot[color=gray, thick] table[x index =0, y index =3] {plot_data/Tx_dBm_vs_Rx_SNIR_dB_Rx_-97dBm_IIP2_60dBm_no_DC_cluster_Tx_reg_with_sigma_1_-1_1_tap_for_Q_IMD2.dat};
\addlegendentry{R-IM2RLS, $\bm{L} = 3e-7 \, \text{upperbidiag}\left(1,-1\right)$}


\addplot[color=black, thick, style = densely dashed] table[x index =0, y index =3] {plot_data/Tx_dBm_vs_Rx_SNIR_dB_Rx_-97dBm_IIP2_60dBm_no_DC_cluster_Tx_smooth_matrix_L_1_tap_for_Q_IMD2.dat};
\addlegendentry{R-IM2RLS using \eqref{eq:smoothing_matrix_L} and $\sigma = 3e-7$}

\addplot[color=gray, thick, gray, densely dotted] table[x index =0, y index =3] {plot_data/Tx_dBm_vs_Rx_SNIR_dB_Rx_-97dBm_IIP2_60dBm_no_DC_cluster_Tx_IM2LMS_mu_1_200_gamma_0_001.dat};
\addlegendentry{IM2LMS \cite{Gebhard2017} with $\mu=0.005, \gamma=0.001$}

\addplot[color=red, line width=1] coordinates{(0,3) (23,3)};

\draw[line width=1, color = red, ->] (21,2.581) -- (21,2.983);
\node[scale=\textsizescale, color = red] at (18,2.675) {SINR};
\node[scale=\textsizescale, color = red] at (18,2.6) {improvement};
\node[scale=\textsizescale, color = red] at (16,3.1) {Rx SNR\,=\,3\,dB};
\end{axis} 

\end{tikzpicture}
\caption{Improvement of the Rx \ac{SINR} at different transmitter power levels and an \ac{Rx} mixer \ac{IIP2} of +60\,dBm. The algorithms are using the DC-filtered receive signal, and the R-IM2RLS/IM2LMS algorithms are using the DC-notch filter to remove the DC. The wanted signal at the antenna has the power \mbox{$P_{\text{Rx}}$\,=\,-97\,dBm} and a SNR of 3\,dB.}
\label{fig:SINR_algo_clustered_Tx}
\end{figure}
The IM2LMS uses the step-size $\mu=0.005$, the regularization parameter $\gamma = 0.001$, and the initial coefficient vector \mbox{$\hat{\ve{w}}_{\text{I}}[-1]=[1e-4,0,0,...,0]^T$}. The \mbox{Q-path} IMD2 replica is estimated by a linear normalized \mbox{1-tap} LMS which uses the \mbox{I-path} IMD2 replica estimated by the IM2LMS as reference input. The normalized \mbox{1-tap} LMS uses a step-size of 1, the regularization parameter is set to 1e-7 and the initial coefficient is set to zero.  The value of the step-size is set to the best compromise between steady-state cancellation and convergence time. The convergence of the algorithms is compared using the ensemble \ac{NMSE}, and the steady-state cancellation by the \ac{SINR}. The \ac{SINR} improvement of the \ac{Rx} signal for the different algorithms and regularizations is depicted in Fig.~\ref{fig:SINR_algo_clustered_Tx}.
%
%
%
\begin{figure}[!ht]
\centering
\begin{tikzpicture}
\begin{axis}[
 width=0.6*\columnwidth, 
 height = .45\columnwidth,
 legend style={font=\scriptsize}, 
 xlabel= {LTE10 slots with OSF=2}, 
 ylabel= {NMSE [dB]},
 xmin = 0,
 xmax = 153600,
 ymin = -25,
 ymax = 5,
 xtick={0,30720,61440,92160,122880,153600},
 xticklabels={0,2,4,6,8,10},
 scaled x ticks = false,
 grid,
]


\addplot[color=black, thick] table[x index =0, y index =1] {plot_data/cancellation_performance_smoothed_MA500_IIP2_60dBm_no_DC_clustered_Tx_reg_sigma_I_1_tap_for_Q_IM2.dat};
\addlegendentry{R-IM2RLS, $\bm{L}=3e-7\,\bm{I}$}


\addplot[color=gray, thick] table[x index =0, y index =1] {plot_data/cancellation_performance_smoothed_MA500_IIP2_60dBm_no_DC_clustered_Tx_reg_sigma_1_-1_1_tap_for_Q_IM2.dat};
\addlegendentry{R-IM2RLS, $\bm{L}=3e-7\, \text{upperbidiag}\left(1,-1\right)$}



\addplot[color=black, thick, style = densely dashed] table[x index =0, y index =1] {plot_data/cancellation_performance_smoothed_MA500_IIP2_60dBm_no_DC_clustered_Tx_smooth_matrix_L_1_tap_for_Q_IM2.dat};
\addlegendentry{R-IM2RLS using \eqref{eq:smoothing_matrix_L} and $\sigma = 3e-7$}

\addplot[color=gray, thick, style = densely dotted] table[x index =0, y index =1] {plot_data/cancellation_performance_smoothed_MA500_IIP2_60dBm_no_DC_clustered_Tx_IM2LMS_mu_1_200_gamma_0_001_w0_1e-4.dat};
\addlegendentry{IM2LMS \cite{Gebhard2017} with  $\mu=0.005, \gamma=0.001$}

\end{axis} 
\end{tikzpicture}
\caption{Convergence of the R-IM2RLS with different regularization matrices and the IM2LMS algorithm at the transmit power level of $P_{\text{Tx}}=23\,\text{dBm}$. The algorithms are using the DC-notch filter to suppress the DC.}
\label{fig:convergence_clustered_Tx}
\end{figure}
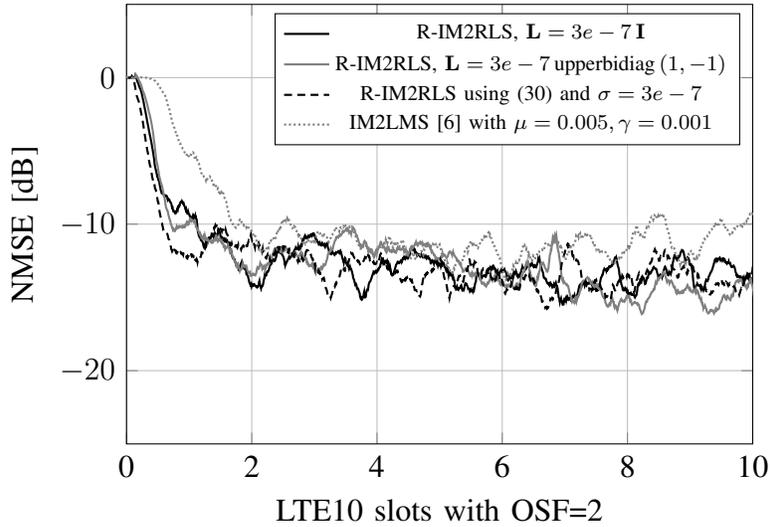
The convergence behaviour of the algorithms is depicted in Fig.~\ref{fig:convergence_clustered_Tx}. The \mbox{R-IM2RLS} shows a faster initial convergence than the IM2LMS algorithm which takes about twice as long to reach an \ac{NMSE} of -10\,dB. 
The evolution of the condition number of \mbox{$\tilde{\bm{R}}'[n] = \bm{P}[n]^{-1}$} is illustrated in Fig. \ref{fig:Rxx_cond_number}.
%
%
%
%
%
%
%
%
%
The condition number of $\tilde{\bm{R}}$ estimated by the \mbox{IM2RLS} without regularization drastically increases up to values between $10^7$ and $10^8$. In contrast to that, the condition number of $\tilde{\bm{R}}'$ estimated by the \mbox{R-IM2RLS} with different regularization matrices $\bm{L}$ stays below 400 for the specific clustered Tx example. The achieved \ac{IIP2} after the digital \ac{IMD2} cancellation is summarized in Table \ref{table:IIP2_improvement_clustered_Tx}. The \mbox{R-IM2RLS} and IM2LMS algorithms are improving the \ac{IIP2} from 60\,dBm to about 77\,dBm and 73\,dBm, respectively.
\begin{table}[ht]
\centering
		\caption{IIP2 improvement by digital cancellation for the clustered Tx signal}
    \begin{tabular}{ | c | c |}
    \hline
    Algorithm           			 & IIP2 after canc. \\ \hline \hline							
    R-IM2RLS, $\bm{L}=3e-7 \, \bm{I}$ 			   	 &		77.2\,dBm	        \\ \hline
		R-IM2RLS, $\bm{L}=3e-7 \,\text{upperbidiag}\left(1,-1\right)  $			   	 &		76.5\,dBm	        \\ \hline
		R-IM2RLS using \eqref{eq:smoothing_matrix_L} and $\sigma = 3e-7$			   	 &		76.4\,dBm	        \\ \hline
		IM2LMS  			   	   &		73\,dBm	        \\ \hline
    \end{tabular}
	\label{table:IIP2_improvement_clustered_Tx}
\end{table}
\section{Verification of the derived algorithm with measurement data} \label{sec:measurements}
\begin{figure}[!ht]
\centering
\begin{tikzpicture}
\begin{axis}[
 width=0.95*\columnwidth, 
 height = .5\columnwidth,
 legend style={font=\scriptsize}, 
 xlabel= {LTE10 slots with OSF=2}, 
 ylabel= {$\left|w_i\right|$},
 xmin = 0,
 xmax = 184320,
 xtick={0,30720,61440,92160,122880,153600,184320},  
 xticklabels={0,2,4,6,8,10,12},
 scaled x ticks = false,
 grid]

\addplot[color=black, thick] table[x index =0, y index =1] {plot_data/coeff_measured_ISP_IMD2_data.dat};
\addlegendentry{$\left|w_i\right|$}
\addplot[color=black, thick] table[x index =0, y index =2] {plot_data/coeff_measured_ISP_IMD2_data.dat};
\addplot[color=black, thick] table[x index =0, y index =3] {plot_data/coeff_measured_ISP_IMD2_data.dat};
\addplot[color=black, thick] table[x index =0, y index =4] {plot_data/coeff_measured_ISP_IMD2_data.dat};
\addplot[color=black, thick] table[x index =0, y index =5] {plot_data/coeff_measured_ISP_IMD2_data.dat};
\addplot[color=black, thick] table[x index =0, y index =6] {plot_data/coeff_measured_ISP_IMD2_data.dat};
\addplot[color=black, thick] table[x index =0, y index =7] {plot_data/coeff_measured_ISP_IMD2_data.dat};
\addplot[color=black, thick] table[x index =0, y index =8] {plot_data/coeff_measured_ISP_IMD2_data.dat};
\addplot[color=black, thick] table[x index =0, y index =9] {plot_data/coeff_measured_ISP_IMD2_data.dat};
\addplot[color=black, thick] table[x index =0, y index =10] {plot_data/coeff_measured_ISP_IMD2_data.dat};

\end{axis} 
\end{tikzpicture}
\caption{Evolution of the estimated coefficients by the R-IM2RLS.}
\label{fig:coeff_measured_data}
\end{figure}
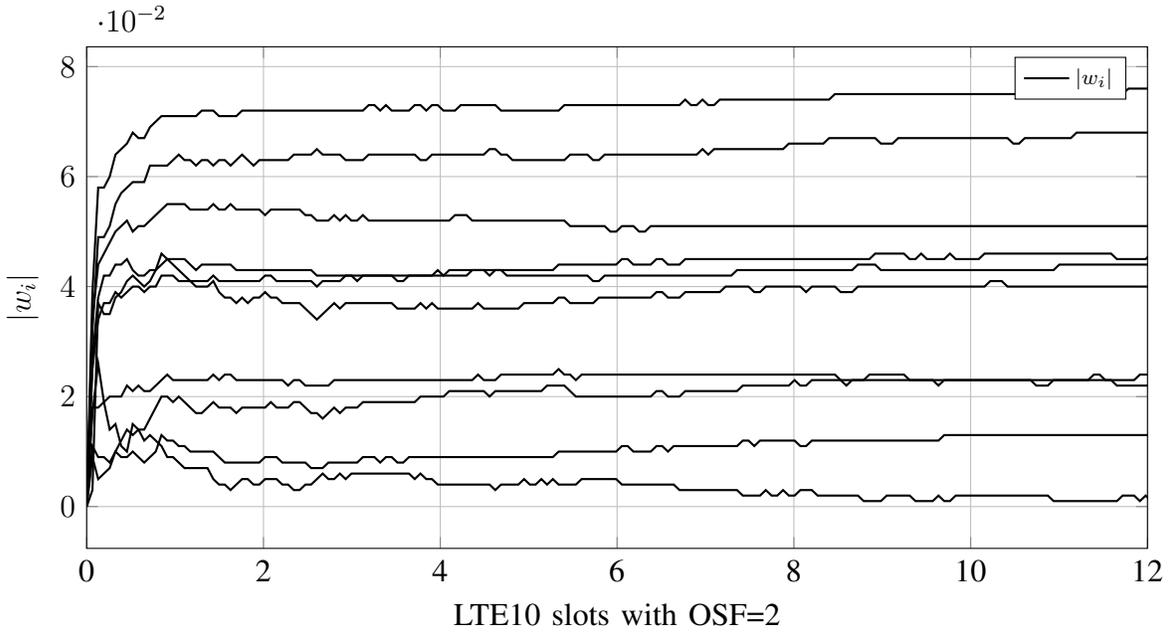

The proposed R-IM2RLS algorithm is evaluated with measurement data and Matlab post-processing. The measurement setup (A) depicted in Fig. \ref{fig:IM2RLS_measurement_setup} includes the \ac{LTE} band 2 duplexer model B8663 from TDK, the \ac{LNA} ZX60-2534MA+ with 41.3\,dB gain and 2.6\,dB \ac{NF} and the ZAM-42 Level 7 mixer which has 25\,dB \mbox{RF-to-LO} terminal isolation. The measurement is carried out for the I-path mixer and a full allocated LTE-A transmit signal with 10\,MHz bandwidth, QPSK modulation and short cyclic prefix. The transmit frequency is set to $f_{\text{Tx}}=1.855\,\text{GHz}$ and the mixer \ac{LO} frequency is $f_{\text{Rx}}=1.935\,\text{GHz}$ (80\,MHz duplexing distance). The LTE transmit signal is generated with the R\&S SMW 200A signal generator (B), and the \ac{TxL} signal which leaks into the receiver with 80\,MHz frequency offset to the \ac{LO} signal is amplified by the \ac{LNA} gain. This amplified \ac{TxL} signal generates the \ac{BB} \ac{IMD2} interference at the output of the \mbox{I-path} mixer which is measured with the real-time oscilloscope \mbox{RTO 1044 (C)}. The \ac{TxL} signal after the \ac{LNA} is measured by the R\&S FSW26 spectrum analyzer (D), and the \ac{LO} signal with 7\,dBm for the ZAM-42 mixer is generated by the R\&S SMB 100A signal generator (E).
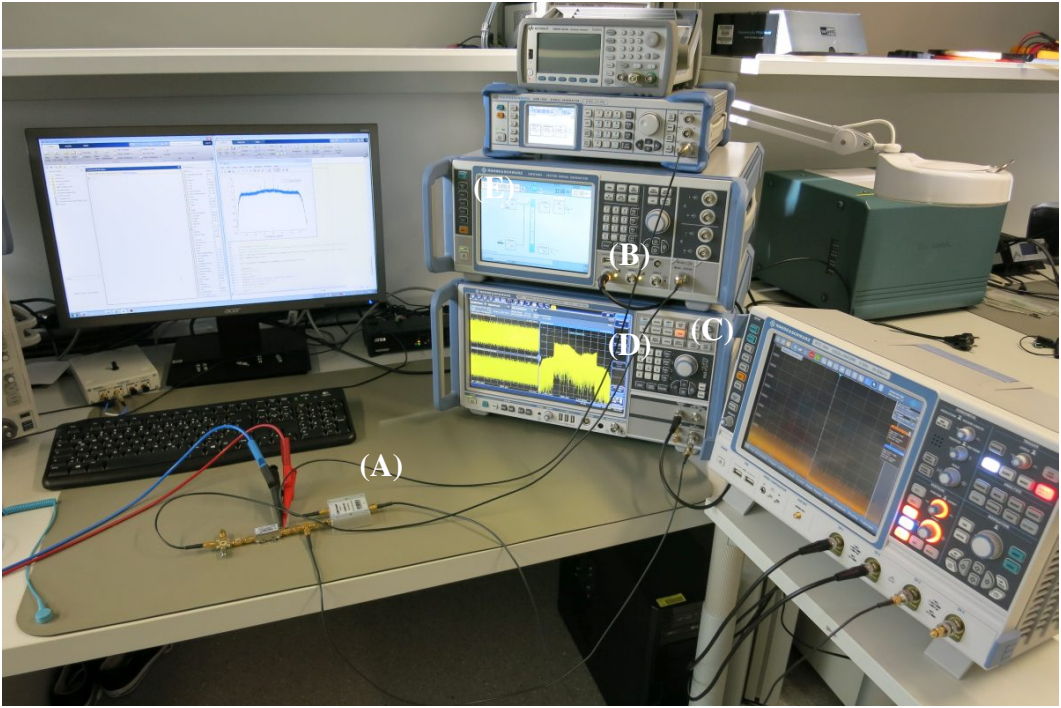
\begin{figure}[ht]
\centering
\begin{tikzpicture}
\pgfmathsetlengthmacro{\textsizescale}{1}
\node (img) at (0,0) {\includegraphics[width=0.8\linewidth,trim=1cm 0.7cm 0cm 0cm]{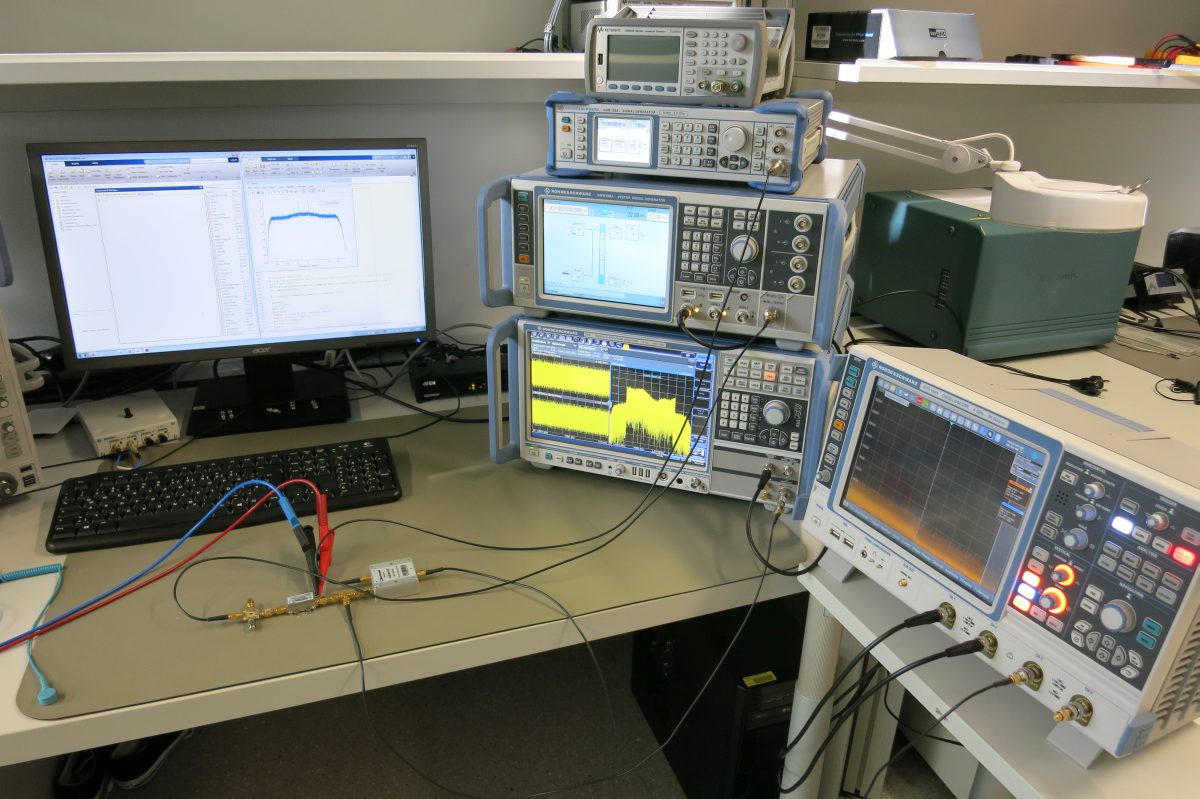}};

\node[scale=\textsizescale, color = white] at (-0.9cm,1.9cm) {\textbf{(E)}};
\node[scale=\textsizescale, color = white] at (0.9cm,1cm) {\textbf{(B)}};
\node[scale=\textsizescale, color = white] at (2cm,0cm) {\textbf{(C)}};
\node[scale=\textsizescale, color = white] at (-2.4cm,-1.8cm) {\textbf{(A)}};
\node[scale=\textsizescale, color = white] at (0.9cm,-0.2cm) {\textbf{(D)}};

\end{tikzpicture}
\caption{
Measurement setup including the DUT (A) with the \ac{LNA} ZX60-2534MA+, the mixer ZAM-42 from Mini Circuits and the \ac{LTE} band 2 duplexer B8663. The signal generator R\&S SMW 200A (B) generates the \ac{LTE} transmit signal and the R\&S real-time oscilloscope RTO 1044 (C) is used to measure the \ac{BB} signal after the mixer. The R\&S FSW26 spectrum analyzer (D) is used to measure the \ac{TxL} signal, and the signal generator R\&S SMB 100A (E) generates the mixer \ac{LO} signal. 
}
\label{fig:IM2RLS_measurement_setup}
\end{figure}
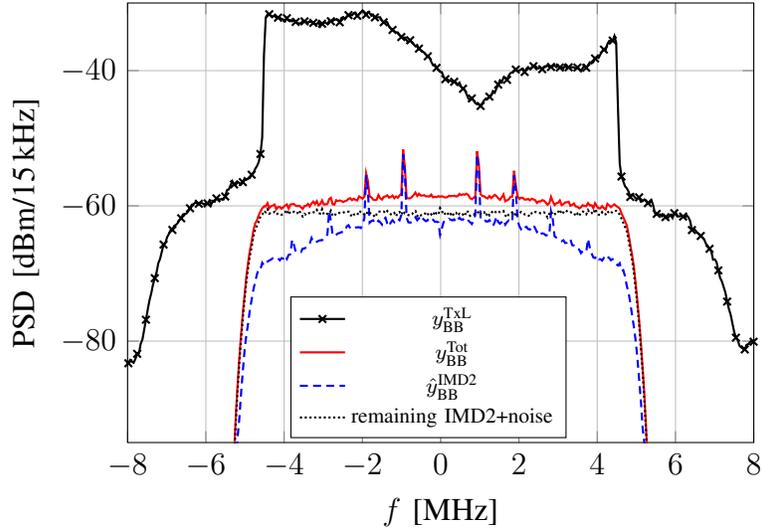
\begin{figure}[ht]
\centering
\begin{tikzpicture}
	\begin{axis}[
	    legend style={font=\scriptsize},
			width=0.6*\columnwidth, 
			height = .45\columnwidth,
			legend columns = {1},
			grid, 
			xlabel={$f$ [MHz]}, 
			ylabel={PSD [dBm/15\,kHz]},
			xmin = -8,
			xmax = 8,
			ymin = -95,
			ymax = -30,
			ytick={-200,-180,-160,-140,-120,...,-60,-40,-20},
			legend style={at={(0.26,0.01)},anchor=south west}
			]

				\addplot[color=black, thick, mark = x, mark repeat={5}] table[x index =0, y index =1] {plot_data/spectrum_measured_ISP_Lab_IMD2_data.dat};
			\addlegendentry{$y_{\text{BB}}^{\text{TxL}}$}
			
			\addplot[color=red, thick] table[x index =0, y index =2] {plot_data/spectrum_measured_ISP_Lab_IMD2_data.dat};
			\addlegendentry{$y_{\text{BB}}^{\text{Tot}}$}
			
			\addplot[color=blue, thick, style = densely dashed] table[x index =0, y index =3] {plot_data/spectrum_measured_ISP_Lab_IMD2_data.dat};
			\addlegendentry{$\hat{y}_{\text{BB}}^{\text{IMD2}}$}
			
			\addplot[color=black, thick, style = densely dotted] table[x index =0, y index =4] {plot_data/spectrum_measured_ISP_Lab_IMD2_data.dat};
			\addlegendentry{remaining IMD2+noise}

  \end{axis}
\end{tikzpicture}
\caption{Spectrum of the measured \ac{TxL} signal $y_{\text{BB}}^{\text{TxL}}$ and the receive signal $y_{\text{BB}}^{\text{Tot}}$ including noise and the \ac{IMD2} interference. The \ac{BB} equivalent \ac{TxL} signal shows a strong frequency selectivity. Also shown are the spectrum of the estimated IMD2 replica $\hat{y}_{\text{BB}}^{\text{IMD2}}$ and the remaining IMD2 and noise after the cancellation.  
}
\label{fig:measured_ISP_IMD2_data}
\end{figure}
The transmit power is set to $P_\text{RF}^{\text{Tx}} =19.3\,\text{dBm}$, which leads in combination with the duplexer attenuation of 67.6\,dB (at $f_{\text{Tx}}=1.855\,\text{GHz}$) and the \ac{LNA} gain of 41.3\,dB to the typical \ac{TxL} signal power of \mbox{$P_{\text{RF}}^{\text{TxL}}=19.3\,\text{dBm}-67.6\,\text{dB}+41.3\,\text{dB} = -7\,\text{dBm}$}.
The measured \mbox{I-path} mixer \ac{BB} output data stream and the complex valued \ac{BB} transmit samples are used for the Matlab post-processing. The spectrum of the signals before and after digital cancellation with the \mbox{R-IM2RLS} using a Tikhonov regularization and the parameters \mbox{$\bm{P}[-1]=10 \bm{I}$},  \mbox{$\lambda = 0.99999$} and \mbox{$\bm{L} = 1e-5 \bm{I}$} are depicted in Fig.~\ref{fig:measured_ISP_IMD2_data}. The Matlab post-cancellation showed that 10 taps were sufficient to cancel the \ac{IMD2} interference by 2.2\,dB down to the noise floor. The coefficient vector was initialized with \mbox{$\ve{w}_{\text{I}}[-1]=[1e-6,0,0,...,0]^T$}, and the convergence of the coefficients is shown in Fig.~\ref{fig:coeff_measured_data} which indicates that the coefficients converged after about 5 LTE symbols. 
%

\section{Conclusion}
This paper presented a novel nonlinear \ac{RLS} type adaptive filter (IM2RLS) and its robust version (R-IM2RLS) for the digital \ac{IMD2} self-interference cancellation in \ac{LTE} \ac{FDD} \ac{RF} transceivers. The R-IM2RLS provides stability and numerical tractability for highly correlated transmit signals which may result in an ill-conditioned autocorrelation matrix. The proposed \mbox{R-IM2RLS} is able to cancel the \ac{IMD2} interference generated by a highly frequency-selective \ac{Tx} leakage signal, and its performance is evaluated with different regularization matrices. Typical \ac{RF} receivers use a DC cancellation to prevent the \ac{ADC} form saturation and a \ac{CSF} to limit the signal bandwidth. Therefore the \ac{IMD2} interference which is generated by the second-order nonlinearity in the mixer is DC filtered and its bandwidth is reduced to the \ac{LTE} signal bandwidth. Consequently, the adaptive filter needs to provide a DC-filtered in-band \ac{IMD2} replica. This contribution shows that the \mbox{IM2RLS/R-IM2RLS} adaptive filter is able to reproduce the in-band \ac{IMD2} interference without DC by including the \ac{CSF} and a DC-notch filter within the algorithm. It is shown, that the proposed algorithm may have multiple solutions of the estimated coefficient vector because of the envelope-squaring nature of the \ac{IMD2} interference. The algorithm converges within a view \ac{LTE} symbols and the steady-state \ac{Rx} \ac{SNR} degradation by the \ac{IMD2} self-interference in case of an multi-cluster transmit signal is improved in simulation from 1\,dB to less than 0.05\,dB. The performance of the R-IM2RLS is proved in an \ac{LTE} measurement scenario with discrete RF components. The \ac{IMD2} interference in the received signal is canceled to the noise floor and a convergence of the coefficients within 5 LTE symbols is achieved. 
\section*{Acknowledgment}
The authors wish to acknowledge DMCE GmbH \& Co KG, an Intel subsidiary for supporting this work carried out at the Christian Doppler Laboratory for Digitally Assisted RF Transceivers for Future Mobile Communications.
The financial support by the Austrian Federal Ministry of Science, Research and Economy and the National Foundation for Research, Technology and Development is gratefully acknowledged.
%

\bibliographystyle{references/IEEEtran_isp}
\bibliography{references/global_bibliography}


\end{document}